\documentclass{aa}  

\usepackage{graphicx}
\usepackage{xcolor}
\usepackage{txfonts}

\usepackage{hyperref}

\newcommand{\objname}{PS16dtm}
\newcommand{\hostname}{SDSS\,J015804.75$-$005221.8}

\newcommand{\todo}{\ifmmode \text{\color{red}\Huge{\(\bullet\)}} \else {\color{red}{\Huge$\bullet$}}\fi}
\newcommand{\tido}{\ifmmode {{\color{red}\bullet}} \else {\color{red}$\bullet$}\fi}

\newcommand{  \Halpha   }{\ifmmode {\rm H}\alpha \else H$\alpha$\fi}

\newcommand{  \ha       }{\Halpha}
\newcommand{  \Hbeta    }{\ifmmode {\rm H}\beta \else H$\beta$\fi}

\newcommand{  \hb       }{\Hbeta}

\def\HeIIop{\ifmmode \textrm{He}\,\textsc{ii}\,\lambda4686 \else He\,\textsc{ii}\,$\lambda4686$\fi}

\newcommand{  \Lop      }{\ifmmode L_{5100} \else $L_{5100}$\fi}

\newcommand{ \Lha   }{\ifmmode L_{\ha} \else $L_{\ha}$\fi}

\def\oiii{\ifmmode \left[{\rm O}\,\textsc{iii}\right]\,\lambda5007 \else [O\,{\sc iii}]\,$\lambda5007$\fi}

\newcommand{  \feii     }{\ifmmode {\rm Fe}\,\textsc{ii} \else Fe\,\textsc{ii}\fi}

\newcommand{\pc}	{\ifmmode {\rm pc} \else pc\fi}
\newcommand{\kpc}	{\ifmmode {\rm kpc} \else kpc\fi}
\newcommand{\ld}	{\ifmmode {\rm l.d.} \else l.d.\fi}
\newcommand{\kms}	{\ifmmode {\rm km\,s}^{-1} \else km\,s$^{-1}$\fi}
\newcommand{\cc}	{\ifmmode {\rm cm}^{-3}    \else cm$^{-3}$\fi}
\newcommand{\cmii}	{\ifmmode {\rm cm}^{-2}    \else cm$^{-2}$\fi}
\newcommand{\ergs}	{\ifmmode {\rm erg\,s}^{-1} \else erg s$^{-1}$\fi}
\newcommand{\ergcms}	{\ifmmode {\rm erg\,cm}^{-2}\,{\rm s}^{-1} \else erg\,cm$^{-2}$\,s$^{-1}$\fi}
\newcommand{\ergcmsA}	{\ifmmode {\rm erg\,cm}^{-2}\,{\rm s}^{-1}\,{\rm\AA}^{-1}
\else erg\,cm$^{-2}$\,s$^{-1}$\,\AA$^{-1}$\fi}
\newcommand{  \ergcmsHz  }{\ifmmode{\rm erg\,cm}^{-2}\,{\rm s}^{-1}\,{\rm Hz}^{-1}
                       \else ergs\,cm$^{-2}$\,s$^{-1}$\,Hz$^{-1}$\fi}
\newcommand{\kev}	{\ifmmode {\rm keV} \else keV\fi}

\def\niii{\ifmmode \textrm{N}\,\textsc{iii} \else N\,\textsc{iii}\fi}
\def\NIII{\ifmmode \textrm{N}\,\textsc{iii}\,\lambda4640 \else N\,\textsc{iii}\,$\lambda4640$\fi}
\def\OIIIbf{\ifmmode \textrm{O}\,\textsc{iii}\,\lambda3133 \else O\,\textsc{iii}\,$\lambda3133$\fi}

\newcommand{  \mbh      }{\ifmmode M_\mathrm{BH} \else $M_\mathrm{BH}$\fi}
\newcommand{  \lmbh     }{\ifmmode \log\left(\mbh/\Msun\right) \else $\log\left(\mbh/\Msun\right)$\fi}

\newcommand{\Msun}{\ifmmode M_{\odot} \else $M_{\odot}$\fi}

\newcommand{  \kbol     }{\ifmmode k_{\rm bol} \else $k_{\rm bol}$\fi}

\newcommand{  \lamLlam  }{\ifmmode \lambda L_{\lambda} \else $\lambda L_{\lambda}$\fi}

\newcommand{\Hubble}	{\ifmmode {\rm km\,s}^{-1}\,{\rm Mpc}^{-1} \else km\,s$^{-1}$\,Mpc$^{-1}$\fi}
\begin{document}

   \title{Tidal disruption of a low-mass star in an active galactic nucleus as the origin of the \objname\,outburst}


\titlerunning{Tidal disruption in an active galactic nucleus}
\authorrunning{M. \'Sniegowska et al.}

\author{Marzena {\'S}niegowska\thanks{sniegowska@mail.asu.cas.cz}\inst{1,2} 
 \and
Bo{\.z}ena Czerny\inst{3} 
\and
Michal Zaja\v{c}ek \inst{4}
\and
Valentina Rosa\inst{1}
\and
Vladim\'{\i}r Karas\inst{1}
\and
Taj Jankovi\v{c} \inst{5}
\and
Tanja Petrushevska\inst{6}
\and
Dragana Ili\'{c}\inst{7,8}
\and
Benny Trakhtenbrot\inst{2}
\and
Petr Kurf\"urst \inst{4}
}

\institute{
Astronomical Institute, Czech Academy of Sciences, Bo\v{c}n\'{i} II 1401, Prague, 14100 Czech Republic \and
 School of Physics and Astronomy, Tel Aviv University, Tel Aviv 69978, Israel
 \and  Center for Theoretical Physics, Polish Academy of Sciences, Al. Lotnik\'ow 32/46, 02-668 Warsaw, Poland
 \and
 Department of Theoretical Physics and Astrophysics, Masaryk University, Kotlá\v{r}ská 267/2, 611 37 Brno, Czech Republic
 \and
 Institute of Physics of the Czech Academy of Sciences, Na Slovance 1999/2,182 21 Praha 8, Prague, Czech Republic
 \and  Center for Astrophysics and Cosmology, University of Nova Gorica, Vipavska 11c, 5270 Ajdov\v{s}\v{c}ina, Slovenia
 \and Department of Astronomy, University of Belgrade - Faculty of Mathematics, Studentski trg 16, 11000 Belgrade, Serbia
 \and  Hamburger Sternwarte, Universitat Hamburg, Gojenbergsweg 112, 21029 Hamburg, Germany \\}

 
  \abstract
   {The event \objname, which occured in the center of the Narrow Line Seyfert 1 (NLS1) galaxy SDSS J015804.75-005221.8 (z = 0.080440), is one of the few candidates for a tidal disruption event in an already-acretting active galactic nucleus (AGN).
   }
   { We aim to shed light on the character of the tidal disruption event in this source since it exhibits 
   unusual peculiarities, such as the double-peak optical/UV light curve and a low blackbody temperature with a lack of X-ray emission.}
   {We perform spectral analysis of the source before and during the event. We model the time evolution of the luminosity profile using a numerical code that describes the viscous evolution of the flow.}
   {From the combined spectral and timing studies, we interpret the event as the disruption of a $\sim 0.3 M_{\odot}$ main-sequence star, or gradual partial disruption of the low-mass giant star. The star is likely on a circular orbit, embedded in the accretion disc. The discussion of the evolution of the star rather suggests that the orbit is counter-rotating. We observe the system at a sufficiently large viewing angle that the actual disruption process is not directly observed. The disrupted star and inner disc are shielded from the observer by a gaseous envelope. Further observations of the system returning to the previous NLS1 state, particularly in the X-ray band, are needed to confirm the proposed scenario and to put constraints on the return to a regular NLS1 state.}
   {}


\maketitle
%
\section{Introduction}
In recent years, rare transient flaring phenomena identified through rapid photometric and spectroscopic variability in galactic nuclei have provided key insight into the physical processes governing the central engines of active galactic nuclei (AGN).
Among the diverse population of SMBH-related transients, the most extensively studied classes are stellar Tidal Disruption Events \cite[TDEs;][]{2012gezari, 2014arcavi, Gezari21_rev}, and Changing-Look AGN \cite[CL AGN;][]{2023riccitrakhtenbrot}, although additional nuclear transient phenomena have also been identified. Tidal disruption events (TDEs) were proposed as early as the mid-1970s as a viable mechanism for fueling SMBHs and sustaining the activity of Seyfert galaxies and quasars \citep{1975Natur.254..295H}.


In the UV–optical band, TDEs are characterized by a bright flare, with a rapid rise in flux (over a few weeks) followed by a more gradual decline \citep{Gezari21_rev,2026arXiv260328380M}. The X-ray emission associated with the formation of a compact accretion disc typically appears with some delay \citep[see e.g.][]{2024SSRv..220...29Z}. The X-ray flux can exhibit quasiperiodic variations due to the Lense-Thirring precession \citep{2024Natur.630..325P} if the angular momentum of the accretion flow is misaligned with the SMBH spin.
Although CL AGN show more variety in the light curve shape in comparison with TDEs, they exhibit overall less dramatic photometric changes compared to TDEs.
From the optical spectroscopic point of view, 
TDEs are also characterized by the presence of an extremely broad ($\gtrsim$ 10000\,\kms) \HeIIop\ \citep{2012gezari, 2014arcavi}, whereas CL AGN exhibit the significant change (or full (dis)appearance) of broad ($\gtrsim$ 2000\,\kms) emission lines. Some of the most prominent flares observed in AGN were suggested to be driven
by a TDE \citep{2015Merloni}, like 1ES\,1927+654 \citep{Trakhtenbrot2019_1ES,Ricci2020_1ES}.
\cite{2025xueguang} suggests that photometric variability of Mrk 1018 \citep[CL AGN explained also by state transitions in][]{noda2018} can be described by the oversimplified double TDEs.
Recent observations reveal even more peculiar cases of flares occurring in previously known AGN. These include AT 2021aeuk, which exhibited multiple (triple) flares and has been interpreted as a repeating partial TDE candidate \citep{bao2024,2025Sun_rpTDE}, and AT 2021loi \citep{Makrygianni2023}, classified as one of the “Bowen Fluorescence Flares” \cite[BFFs;][]{2019Trakhtenbrot}. BFFs are characterized by exceptionally strong and broad ($\approx2000$\,\kms) Bowen emission lines, such as \NIII\ and \OIIIbf, along with the \HeIIop\ line that is directly linked to the Bowen fluorescence mechanism \citep{Bowen1928, Netzer1985}. There are a number of sources classified as Ambiguous Nuclear Transients (ANTs) where it is not clear if we observe a Changing-Look phenomenon in an AGN, or TDE in a non-active or weakly-active galaxy \citep[see e.g.][]{2025wiseman, clark2025}. The interesting class of Extreme Nuclear Transients (disruptions of massive stars by massive black holes, \citealt{hinkle2025}) usually takes place in non-active galaxies, although some may be related to weakly active ones \citep[e.g.][]{onori2022,hoogendam2024}. 
Those events are rare, and their nature deserves a detailed multiwavelength study. However, as time-domain astronomy advances, the number of peculiar sources is growing, and we expect to increase the number of known events with wide-field optical surveys, such as the Vera Rubin Observatory Legacy Survey of Space and Time \citep{2019ApJ...873..111I}, which will be aided by planned UV-domain missions that are especially relevant for nuclear transients \citep[the Ultraviolet Transient Astronomy Satellite -- ULTRASAT, the Quick Ultra-Violet Kilonova surveyor mission -- QUVIK, the Ultraviolet Explorer -- UVEX;][]{2021arXiv211115608K,2024Shvartzvald,2024SSRv..220...11W}. 



In this work, we focus on \hostname\, a rare candidate TDE occurring in a previously known AGN. The event \objname\ was a nuclear transient discovered on 12 August 2016 (MJD = 57612) by the Pan-STARRS Survey for Transients \citep[PSST;][]{2016chambers} in a nearby (z = 0.080440 $\pm$ 0.00001; NED) active galaxy \hostname\ \citep{blanchard2017, 2017ApJ...850...63J, petrushevska2023,2025jiang}.  The flare brightened 2 mag above the host in $\sim $ 50 days and showed an unusual second peak at $\sim $ 100 days after the first detection in UV/optical bands. After approximately 2000 days of monitoring after the outburst, the source returned to near pre-outburst brightness levels, while the mid-infrared luminosity continued to rise slowly \citep{petrushevska2023}.
 This transient also shows strong Fe II emission after the flare 
\citep{petrushevska2023}. The peak bolometric luminosity is estimated as $\sim 10^{44.62}\,{\rm erg\,s^{-1}}$ \citep{petrushevska2023}, which corresponds to the Eddington luminosity of a black hole with a mass of $3.3 \times 10^6 M_{\odot}$.

Before the outburst, the host was classified as a dwarf Seyfert 1 galaxy \citep{greene_2_2007, Xiao2011} with narrow emission lines \citep{Xiao2011} which led to its classification as Narrow Line Seyfert 1 (NLS1). During the outburst, the spectrum of the NLS1 galaxy \citep{blanchard2017} changed,  suggesting that the source indeed reached the Eddington accretion rate. The black hole mass was estimated to be $\log{(M_\mathrm{BH}\,[M_{\odot}])}\simeq 5.9$ \citep{Xiao2011}, which would imply even super-Eddington state during the outburst \citep{blanchard2017}. Emission-line ratio diagnostics applied to the pre-outburst spectrum indicate the presence of AGN activity, although the relatively strong [N II] emission places the source closer to the star-forming (H II-region) locus in standard diagnostic diagrams. The diagnostic diagrams in \cite{petrushevska2023}locate \hostname\ in the area of AGN-starburst composite objects (see Fig. 16 therein). 
The source has been studied in detail in a number of publications \citep{blanchard2017, petrushevska2023,2024ApJ...971..185C,2025jiang}. Some issues remained, including the problem of the temperature of the accreting material \citep{petrushevska2023}. 

In this paper, we present a global model of the time evolution of the source, assuming a TDE origin.
In Section~\ref{sect:before} we reanalyze the state of the AGN nucleus prior to the TDE. In Section~\ref{sect:toy} we use a simple model of viscous spreading of the ring formed due to TDE, focusing on potentially multiple material deposits which can best reproduce the observed lightcurve in \objname. Finding a correspondence between the viscous timescales and the deposit radius requires the knowledge of the material temperature, and with this aim, we reanalyze the broad band Spectral Energy Distribution (SED) of the source in its bright state in Section ~\ref{sect:SED}. We explore the geometry of the event in Section~\ref{sect:geom}, while in Section~\ref{sec:nature-of-dusruption} we focus on the nature of the disrupted object, including the basic dynamics of star-disc interactions and associated timescales.
We discuss additional aspects of TDEs in AGN in Section~\ref{sec:discussion}. Finally, we conclude with Section~\ref{sec:conclusions}.
Throughout this paper, we assume a cosmological model with  $H_0 = 70\,\Hubble$, $\Omega_\Lambda = 0.7$ and $\Omega_{\rm M}=0.3$.






\section{The state of the AGN before the flare}
\label{sect:before}

\begin{figure*}
    \centering

       \includegraphics[scale=0.5]{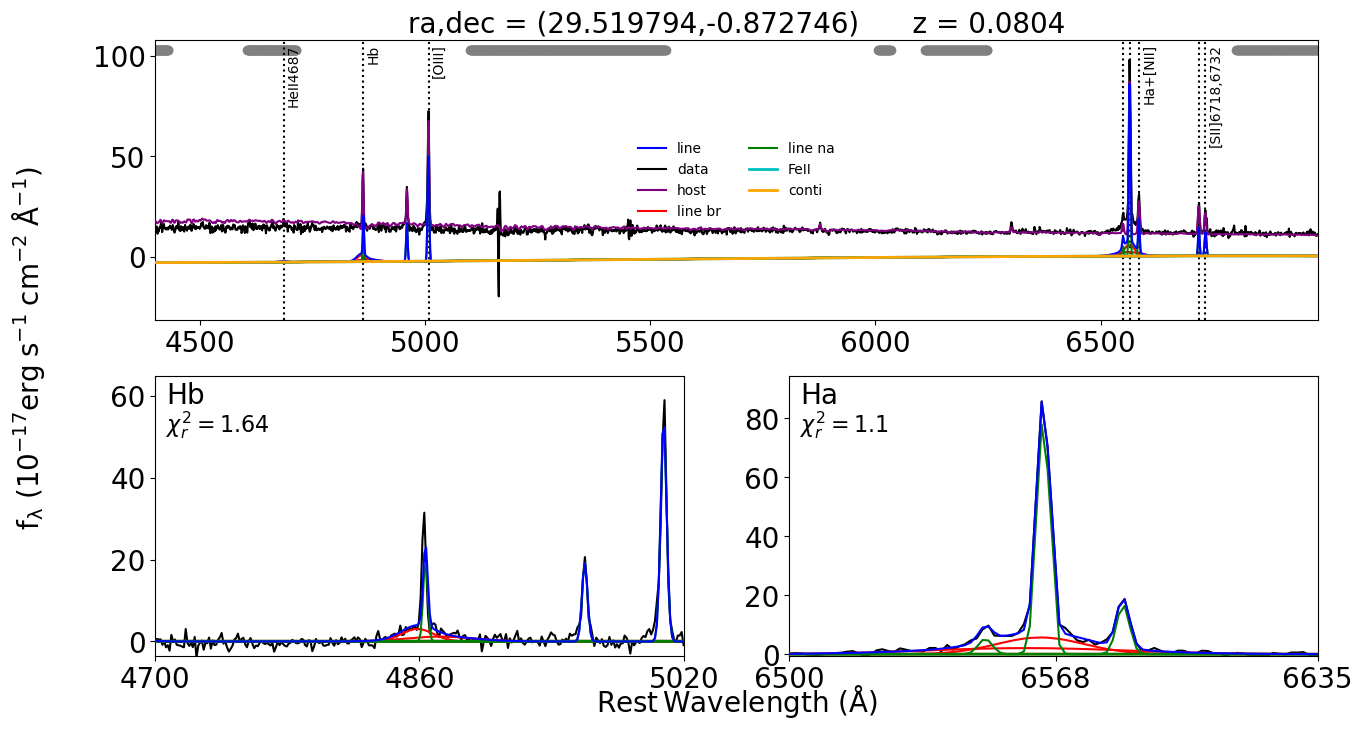}

    \caption{Spectral decomposition of archival (MJD = 52933) SDSS spectrum of \hostname\ taken before the \objname\ event detection (MJD = 57612).  \textit{Top panel:}
    The overall spectrum (in black) and best-fit model (in blue). The model components include: continuum emission (orange), fitted host-galaxy contribution (purple), iron pseudo-continuum (cyan), and broad \& narrow components of various emission lines (red and green, respectively). The gray thick horizontal lines near the top mark the spectral windows used for the continuum estimation.
    \textit{Bottom panel:} zoom-in views of the H$\beta$ and H$\alpha$ spectral regions.}
    \label{fig:sdss-spectrum}
\end{figure*}

Since properties of the \objname\ are not typical for a TDE, we first reanalyze the available spectrum of the host before the flare, to better constrain the geometry prior to the event. Bright AGN (above a few percent of the Eddington ratio) typically have cold standard accretion discs approaching the Innermost Stable Circular Orbit (ISCO), while fainter AGN show a hot advection-dominated accretion flow (ADAF) at tens or hundreds of Schwarzschild radii ($R_{\rm{}Schw}$). The SDSS spectrum of \objname\ has been analyzed in detail by \citet{blanchard2017} and \citet{petrushevska2023}; nevertheless, we revisit this analysis to better constrain the pre-outburst state of the source.  The spectrum obtained on MJD = 52933, shown in Fig. \ref{fig:sdss-spectrum}, is strongly dominated by narrow lines, but a weak broad component in the H$\alpha$ and H$\beta$ region is clearly present, indicating that the source should be classified as Seyfert 1.8 \citep{osterbrock1981}. However, in this classification the \ha/\hb\ ratio, or Balmer decrement, plays an important role as the line weakness is related to the dust extinction. The distinction between Seyfert 1.8 and 1.9 objects is therefore usually based on the detectability and relative strength of broad \hb, often assessed with respect to \oiii. Seyfert 1.8 galaxies retain a weak or marginal broad \hb\ component together with broad \ha, while Seyfert 1.9 galaxies show broad \ha, but no detectable broad \hb \citep{1992winkler}.


To decompose the spectrum, we used {\sc PyQSOFit} \citep{pyqsofit}. We model the accretion-powered continuum emission with a power-law. The host-galaxy contribution is described using a principal component analysis approach \citep{2004AJyip, 2004yip} implemented in {\sc PyQSOFit}. To model narrow and broad components of emission lines, we used Gaussian profiles. For a better fit, we used 2 Gaussians for modelling broad components, and we report the FWHM of combined profiles.
The narrow component of \hb\ and \ha\ have $FWHM=200\,\kms$, while  \oiii\  has $FWHM=250\,\kms$.
We checked the SDSS spectral resolution directly from the FITS spectrum and evaluated the instrumental width at the observed wavelength of each narrow emission line mentioned above in order to assess whether the narrow components are resolved. We find instrumental FWHM values of 145, 141, and 132 \kms\  for \hb, \oiii, and \ha, respectively. After correcting the measured widths in quadrature for instrumental broadening, the observed FWHM values of 204, 259, and 198 \kms, 
 correspond to intrinsic widths of 144, 217, and 147 \kms, respectively. Thus, the narrow lines are resolved in the SDSS spectrum, although the instrumental correction is not negligible. The narrow components are used here only for the purpose of spectral decomposition.

In Table \ref{tab:sdss-measurements}, we present measurements of the broad components of H$\beta$ and H$\alpha$ lines from the SDSS archival spectrum.
We confirm that the FWHM of H$\alpha$ is narrow, 1400$\pm$150 \kms, comparable to the determination of 
\citet{greene_2_2007} equal to 1680 \kms. This measurement in our fit is affected by the strong contribution of narrow lines in H$\alpha$ range. FWHM of H$\beta$ from our fit is 1680$\pm$200\kms. For the broad component FWHM of some of the emission lines, the effect of the instrumental spectral resolution is negligible compared to the measurement uncertainties. Thus, we do not include this correction. Using our independent measurements of the narrow line ratios log([O III] $\lambda$5007/H$\beta$) = 0.54, log([N II] $\lambda$6583/H$\alpha$) = -0.69, and log([S II] $\lambda\lambda$6717/H$\alpha$) = -0.74, we confirm the placement of \hostname\ in the area of AGN-starburst composite objects, as stated in \cite[][see Fig. 16]{petrushevska2023}.


Continuum of the fit is dominated by the host galaxy, thus we decided not to rely on \Lop\ (= 10$^{43.02}$ \ergs) from our fit in estimating source properties, like \mbh.
To show that this part of spectrum is indeed host-dominated, we also estimate the host luminosity using synthetic host photometry from SWIFT (V-band with effective wavelengths of 5468 \AA, which is the most relevant to 5100 \AA) for \hostname\ \citep{2021hinkle}, which is 18.43 mag, and we obtain a luminosity of 10$^{43.2}$ \ergs.

To estimate the SMBH mass, we first use the luminosity and width of the broad H$\alpha$ line using the prescription of \citet{MR22}, and we obtain $\log(\mbh/\Msun) =6.0$, which is in agreement with values from the previous studies \citep{Xiao2011}.
From our fitting, we obtain $L_{H_{\alpha}} = 5 \times 10^{40}\,{\rm erg\,s^{-1}}$ and we use this number to estimate \Lop. 
To estimate  $L_{BOL}$, we used relation for $L_{H\alpha} - \Lop$ from \cite{2005greene}
\begin{equation}
L_{H_{\alpha}} = (5.25 \pm 0.02) \times 10^{42} 
\left( \frac{\Lop}{10^{44}~{\rm erg~s^{-1}}}\right)^{(1.157 \pm
                          0.005)}~{\rm erg~s^{-1}} 
\end{equation}

\Lop\ from this calculation is $1.9\times 10^{42} \ergs$, which is 8 times lower than \Lop\ from Swift (V-band) and 5 times lower than \Lop\ from our optical spectral fitting. Combining with \kbol\ = 8.8 for $\Lop$ from \cite{marconi2004}, we obtain $L_{BOL} = 1.7 \times 10^{43} \ergs$. The bolometric (optical/UV) luminosity estimated in this way is an order of magnitude higher than the X-ray luminosity determined to be $L_X=(1.2\pm 0.5) \times 10^{42} \ergs$. This is typical for sources emitting well below the Eddington rate. 
Using these computations, we estimate the Eddington ratio to 13\%.
The ratio of \ha\ to \hb\ (broad component) in the SDSS data is 2.17 $\pm$ 0.20, which is lower than the widely adopted value for the dustless Case B recombination, for which this ratio is higher than 2.5 \citep{1995MNRAS.272...41S}. The low ratio may be caused by the high-accretion nature of the host, lower temperature and higher densities of the BLR gas \citep[i.e.][]{2009ferland, 2012ilic}.

\begin{table*}
\center
\caption{Preflare spectral measurements of \hostname.}
\label{tab:sdss-measurements}

\begin{tabular}{llrll}
\hline
\hline
Quantity & units & value  &    reference \\
\hline
FWHM $H\beta$  & \kms & 1680 $\pm$ 200 & {\rm fit of SDSS spectrum} \\
Flux $H\beta$  & $10^{-17}\,\ergcms$ & 190 $\pm$ 10 & {\rm fit of SDSS spectrum} \\
L $H\beta$ & \ergs & 3 $\times 10^{40}$ \\
 FWHM $H\alpha$ & \kms & 1400 $\pm$ 150 & {\rm fit of SDSS spectrum} \\
Flux $H\alpha$ &  $10^{-17}\,\ergcms$ & 330 $\pm$ 20 &{\rm fit of SDSS spectrum} \\
L  $H\alpha$ & \ergs & 5 $\times 10^{40}$ \\
$\log(M_{\rm BH})$ & $M_\odot$ &    6.0   &  from FWHM and luminosity of $H\alpha$\\
\Lop & \ergs & 43.02$\pm$ 0.01 &  this paper, from SDSS spectrum, no starlight subtraction\\
\Lop & \ergs & 42.28$\pm$ 0.01 &  this paper, from \cite{2005greene} relation \\

\hline

\end{tabular}
\end{table*}

The spectrum obtained 1868 days after the outburst by \citet{petrushevska2023} still shows a much stronger broad H$\alpha$ component (with broad-to-total faction between Broad component and the whole line profile 0.73) in comparison to the pre-outburst SDSS spectrum with fraction 0.54.\citet{petrushevska2023} argue that the difference between the two spectra is purely instrumental (XSHOOTER vs SDSS), and that the width of the line $1160 \pm 190$ km s$^{-1}$ is thus suitable for determination of the black hole mass (log(\mbh/\Msun) = $6.07\pm 0.18$). Our value of the FWHM is thus much narrower than typically in Seyfert 1.8 galaxies. On the other hand, it shows some similarity to spectra of LINERs with broad H$\alpha$ line components \citep{marquez2017}, but the [NII] lines in LINERs (after subtraction of the host contribution) are even stronger than in \objname.
We decided to verify the classification of the host galaxy.
Taking NLS1 criteria \citep{1985Osterbrock, 1989goodrich}:
(i) the FWHM of the \hb\ line $<$ 2000 \kms, (ii) the ratio of narrow components \oiii\,/\hb\ $<$ 3; and (iii) unusually strong \feii\ lines.
 FWHM of \hb\ is $<$ 2000 \kms, and 
the ratio of \oiii\,/\hb\ fluxes from our fitting is 1.1, so two criteria are met by \hostname.
However, this source does not have strong \feii\ lines, and this source may be classified as a rare type
of NLS1  \citep{2006zhou, petrushevska2023}
In general, NLS1s are also characterized by high Eddington ratios \citep[i.e.][]{2016Cracco}. For \hostname, this value (= 0.13) is relatively low compared to the sample of NLS1 from SDSS DR12 \citep{2017rakshit}, in which the mean Eddington ratio is 0.25.
\footnote{We compare \hostname\ to sample of objects with good quality
measurements (quality flag = 0), which is 85\% of sources from \cite{2017rakshit}.}

Overall, the source before the outburst is not an extreme accretor as some of the popular NLS1 sources, for example, I Zw I \citep{Sargent1968,1985Osterbrock}. However, it formally satisfies all the criteria for NLS1 source, and the Eddington ratio indicates that the accretion in the source likely proceeded through a standard accretion disc down to the Innermost Stable Circular Orbit (ISCO).
Inner hot flow, like Advection-Dominated Accretion Flow (ADAF), or more generally, with Radiatively-Inefficient Accretion Flow (RIAF), develops typically in sources with the Eddington ratio below a few percent \citep[e.g.,][and the references therein]{Yuan2004,sniegowska2020}. 
Having the estimates of FWHM of \hb\ and 5100\AA\ luminosity, we can determine the distance to the Broad-Line Region (BLR) using two different methods. First, using the virial equation:
\begin{equation}
    \label{eqn:radius_fromvirial_mass}
   {R}_{\rm BLR} = \frac{G   M_\mathrm{BH}}{f\text{FWHM}^2},
\end{equation}
where $G$ is the gravitational constant, $M_\mathrm{BH}$ is the central BH mass, $\text{FWHM}$ is the full width at half maximum of the emission line, and $f$ is the virial factor, which we assumed to be 0.5 based on Figure 12 from \cite{2024gravity}. 
 From this method, we obtain the BLR radius $3.2\times 10^4$ $R_{\rm{}Schw}$. 
Using the radius–luminosity ($R$–$L$) relation calibrated by the GRAVITY Collaboration, which is based on VLTI/GRAVITY spectro-interferometric observations that spatially resolve shifts across broad emission lines and directly model the BLR geometry and kinematics, rather than measuring time lags between continuum and line variability 
 \citep[see][]{2024gravity, 2026gravity}.
 From this $R-L$ relation, using Equation 4 from \cite{2024gravity}
\begin{equation}
    {\rm log} \ (R_{\rm BLR}/ {\rm ld}) = K + \alpha \ {\rm log}(\lambda L_\lambda/10^{44} \ {\rm erg} \ {\rm s}^{-1}).
\end{equation}
with a normalization coefficient $K = 1.69$ and a slope $\alpha = 0.37$, given in \cite{2024gravity},  we obtain the BLR size of $R_{\rm BLR} \sim 11\,{\rm l.d.}$, which is $10^5$ $R_{\rm{}Schw}$. We thus obtain roughly consistent results with both methods.
\section{Time evolution of the TDE}
\label{sect:toy}
We now model the time evolution of the TDE, aiming to reproduce not only the overall decay of \objname\ but also the rise phase and the plateau. In this work, we do not consider debris fallback and assume that the evolution is dominated by accretion-disc physics.
We concentrate on the photometric data as they follow the entire evolution from the early rise through the decay phase. 

\subsection{Toy-model}
\label{sect:toy_model_description}

The analytical description of the propagation of the mass flow rate in the accretion disc is characterized
by the Green function \citep{lyndenbell74}, which was used in the context of  dwarf novae \citep{1989mineshige},
X-ray binaries \citep{2001kotov,zdziarski2009} or activity of Sagittarius A* \citep{2013czerny}. A similar method, also based on a viscous timescale, has been used by \citet{guolo2025} to model the long-term evolution of GSN 069.
We model a single or multiple accretion events following the prescription used by \cite{zdziarski2009} and \cite{2013czerny}, assuming the deposit of matter in a circular orbit within the accretion disc in the equatorial plane of AGN. Parameters included in this model are the viscous timescale and the mass of the star.
If the infalling star reaches critical distance $R_0$ from the black hole, it is disrupted at $t_0$, and forms a ring which diffuses on the viscous timescale \citep{2011MNRAS.418..276S,2020MNRAS.492.5655M}. If not, we can have multiple deposits of the material at different radii. We assume that each deposit is in the form of a ring, which then spreads on a viscous timescale. Part of the material reaches the ISCO, where the material can most efficiently deposit its gravitational energy.  


In principle, we should solve the equations of continuity,
\begin{equation}
\label{eq:sigma}
\centering
 \frac{\partial  \Sigma }{\partial t} = \frac{1}{2 \pi r} \frac{\partial  \dot M }{\partial  r}\,,
\end{equation}
and angular momentum transport,
\begin{equation}
\label{eq:mdot}
\centering
 \dot M = 6 \pi r^{1/2}  \frac{\partial   }{\partial  r} (r^{1/2} \nu \Sigma)\,.
\end{equation}
Here  $r$ is the radius of the accretion disc, $\Sigma(r)$ is the surface density, and  $\dot M(r)$ is the mass local flow rate. The key parameter is the kinematic viscosity, $\nu$. If it is constant, the entire disc evolution is described by Green’s functions \citep[][]{lyndenbell74,1998kato}:
\begin{eqnarray}
\lefteqn{G_\Sigma(r,\tau) \! = \! {2\Sigma_0 |\mu| \xi^{1/\mu-9/2}\over \tau} \exp \! \left[ -{2\mu^2(\xi^{1/\mu}+1) \over \tau }\right] {\rm I}_{|\mu|} \! \left[ 4\mu^2\xi^{1/(2\mu)}\over \tau \right], 
\label{eq:sig1}} \\
\lefteqn{G_{\dot{M}}(r,\tau) \! = \! {\dot M_0 |\mu|\over \tau} \frac{\partial}{\partial \xi} \xi^{1/2} \exp \! \left[ -{2\mu^2(\xi^{1/\mu}+1) \over \tau }\right] {\rm I}_{|\mu|} \! \left[ 4\mu^2\xi^{1/(2\mu)}\over \tau \right].}
\label{eq:mdot1}
\end{eqnarray}
Here $\Sigma_0$ is the initial surface density of the ring in the form of a Dirac delta function at the radius $R_0$, $\xi$ is the dimensionless radius ($r/R_0$, $\tau$ is the dimensionless timescale ($t/\tau_{\rm visc}$, $\tau_{\rm visc}$ in turn is related to the kinematic viscosity through expression $\tau_{\rm visc}= 2 r^2/(3 \nu)$, and $\mu$ is related to the polytropic index. The initial accretion rate is determined by 
\begin{equation}
\dot M_0 = {4 \pi r^2 \Sigma_0 \over \tau_{\rm visc}(R_0)}.
\end{equation}

However, if we only need to obtain the time profile of the source luminosity, we do not need to solve these two equations for all disc radii. Instead, we can use directly the expression for the Green function for the time-dependent accretion rate at the inner radius of the disc derived by \citet{zdziarski2009}:
\begin{equation}
G_{\dot{M}}(\tau) = {(2 \mu^2)^\mu \over \Gamma(\mu)} \tau^{-(1+\mu)}\exp{\left(-{2 \mu^2 \over \tau}\right)}
\end{equation}
where $\Gamma(\mu)$ is the Euler gamma function. This Green function can be used for a single or multiple deposits. For a single deposit, the accretion rate is obtained by multiplying the Green function by $\dot M_{0}$. The viscous time $\tau_{\rm visc}(R_0)$ is effectively a parameter of the solution. Its conversion to actual radius depends on the disc properties and will be discussed later on. In this section, it is just a free parameter of the model, along with the total mass initially deposited in the ring. The evolution of the bolometric luminosity is calculated from the calculated $\dot M$ profile, assuming the standard efficiency of accretion of 0.1 \citep[see e.g][]{1982soltan}.


\subsection{Results}
\label{sec:toy_results}




\begin{figure*}
\centering
      \includegraphics[scale=0.5]{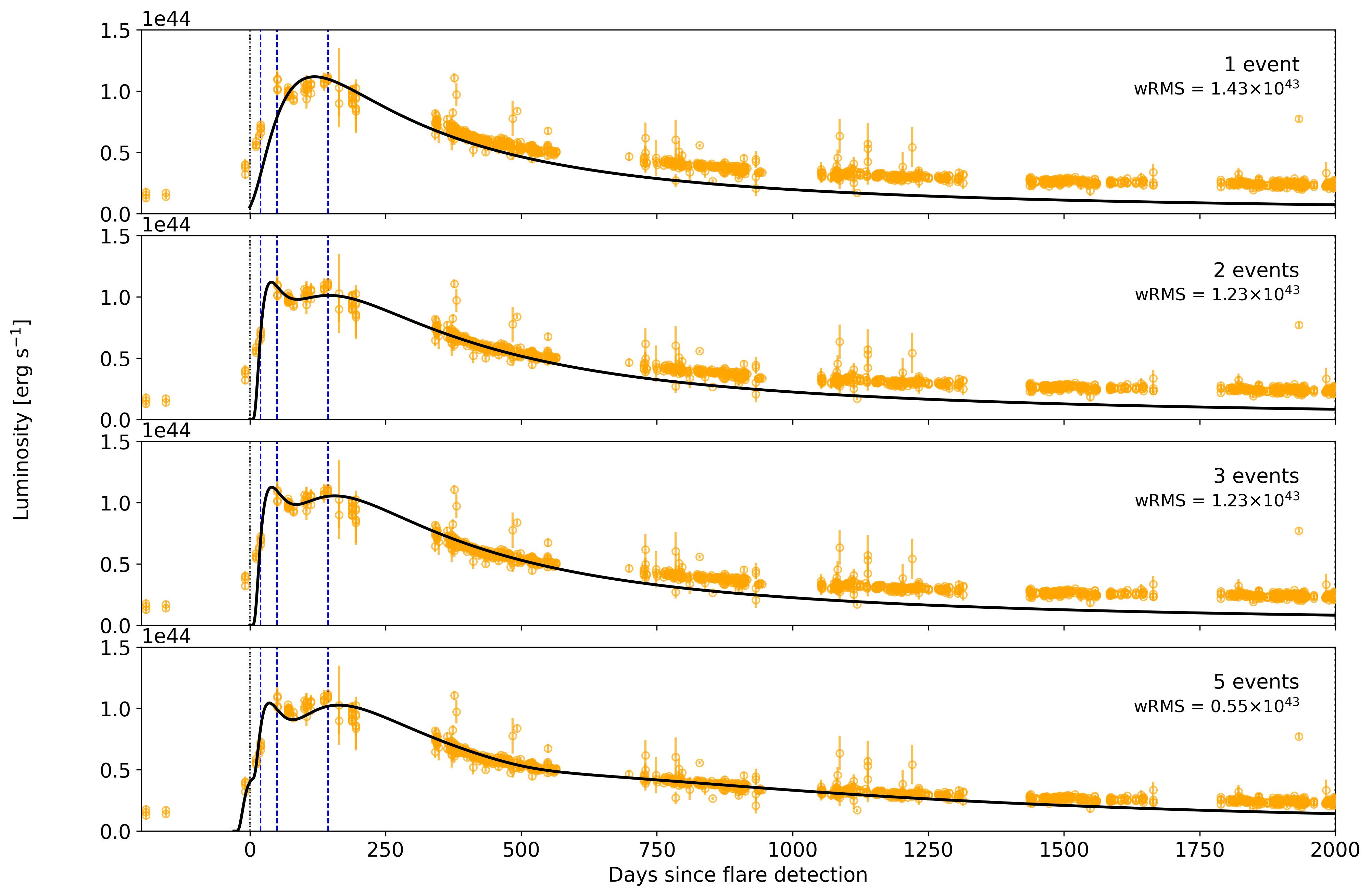}

    \caption{Photometric evolution of \objname\ and lightcurve generated by a discrete accretion event, double, triple, and quintuple accretion events. With orange circles, we mark ATLAS photometry $o$ band with signal-to-noise $>$ 3 and after excluding outliers identified with a robust 5-median absolute deviation cut. The detection of the event MJD = 57612 is defined as t = 0 and marked with a black dotted line. With blue dashed lines, we mark the MJD of Swift observations used in this work for SED fitting (MJD = 57632, 57647, and 57774). Weighted RMS values are listed in each panel for comparison.  }
    \label{fig:lc-event}
\end{figure*}

\begin{figure*}
\centering
      \includegraphics[scale=0.6]{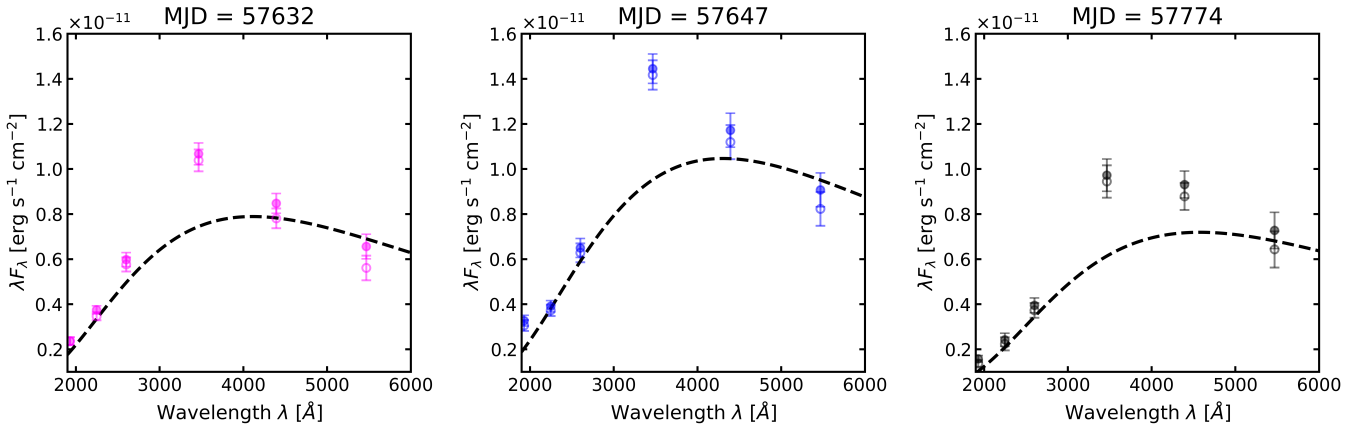}

    \caption{Rest-frame Swift/UVOT observations of \objname\ from MJD = 57632, 57647, and 57774 without (filled circles) or with (open circles) host subtraction. Using dashed lines, we plot black-body SEDs with temperatures of 8900K (left), 8500k (middle), and 8000K (right).
     The host subtraction does not change the shape of the SED significantly. We discuss the visible excess at $\sim 3600 $\AA\, in Section~\ref{sect:SED}.}
    \label{fig:ps16-few-sed}
\end{figure*}

\begin{table*}
\centering
\caption{Values of parameters predicted by fitting the lightcurve with a single or multiple discrete accretion events. Black hole mass was fixed at BH mass $10^6$ \Msun. In the bottom row, we report $wRMS$ for all four cases of model from Figure \ref{fig:lc-event}. }
\label{tab:parameters}
\begin{tabular}{ll|lllll}
\hline
\hline
&   & event 1      & event 2        & event 3 & event 4 & event 5        \\ \hline
Parameter         & unit   & value       & value        & value & value & value        \\ \hline
deposit time$^*$ & year &  -0.1  & - & - & - & -\\
viscous timescale & year & 0.8 & - & - & - & -\\
mass & $M_{\odot}$ & 0.26 & - & - & - & -\\
\hline
deposit time$^*$ & year & 0 & 0.01 & - & - & - \\
viscous timescale & year & 0.2 & 1.1 & - & - & - \\
mass & $M_{\odot}$ & 0.13 & 0.19 & - & - & - \\      
 \hline
 deposit time$^*$ & year & 0 & 0.01 & 0.05 & - & - \\
viscous timescale & year & 0.2 & 0.8 & 1.1 & - & - \\
mass & $M_{\odot}$ & 0.13   &  0.05 & 0.15  & - & - \\ 
\hline
 deposit time$^*$ & year & -0.08 & 0 & 0.01 & 0.05 & 1.0 \\
viscous timescale & year  & 0.2 & 0.2 & 0.8 & 1.1 &  3.1 \\
mass & $M_{\odot}$ & 0.05 & 0.08 &  0.05 & 0.15  & 0.05  \\ 
\hline
$wRMS$ &  &  $1.43 \times 10^{43}$ & $1.23 \times 10^{43}$  & $1.43 \times 10^{43}$ & -  & $0.55 \times 10^{43}$   \\ 
\hline
\hline
\end{tabular}

$^*$ deposit timescale is measured with respect to detection of the event (MJD = 57612).  

\end{table*}

We now use the model described in Section~\ref{sect:toy_model_description} and present our fits in Figure \ref{fig:lc-event}. We focus on modeling the photometry from Asteroid Terrestrial-impact Last Alert System \cite[ATLAS]{2018torny} in the o-band curve\footnote{ MJD = 57398--59983 (2016 January 11 to 2023 February 8)}, as it offers significantly better time coverage for this source in comparison to c-band. We fix the value of the central black hole at $10^6 M_{\odot}$, following \citet{blanchard2017} and \citet{petrushevska2023}.  
We first study the single deposit event (upper panel in Figure \ref{fig:lc-event}), which corresponds to a rapid disruption of the star in a single event.
For this event, we assumed 0.26 $M_{\odot}$, a duration timescale of 0.8 yr, and a deposit time of 0.1 yr before the flare was observed. The fit does not capture well the plateau, and with just one deposit event, it is impossible to mimic the `double hump' during the peak of the event.
\citet{petrushevska2023} also showed that the single power-law model with $\sim t^{-5/3}$ decline does not provide a good fit for the optical observations of \objname. 
In the lower panels, we present fits for 2, 3, and 5 accretion events, respectively.
The parameters of all the fits are listed in Table~\ref{tab:parameters}. 
To evaluate the agreement between our models and the ATLAS o-band photometry, we use the uncertainty-weighted RMS \cite[see, e.g.,][]{2022ApJ...938...83Y} of the residuals corresponding to the period from the flare detection to 2000 days later.
\begin{equation}
\mathrm{wRMS} =
\left[
\frac{\sum_{i} w_i \left(F_i - F_{i,\mathrm{model}}\right)^2}
{\sum_{i} w_i}
\right]^{1/2},
\qquad
w_i = \frac{1}{\sigma_i^2}.
\end{equation}
Here $F_i$ and $\sigma_i$ are the observed fluxes and related uncertainties, respectively, and $F_{i,\mathrm{model}}$ denoted the model value interpolated to the observation time.
Within the models presented in this work, the five-event case provides the best description of the data. We highlight that for all the cases included in the model, the injected mass is fairly similar, varying from 0.26 $M_{\odot}$, for one event, to 0.38 $M_{\odot}$, for five events.
Timescales and masses obtained by this model are not directly connected to radii, because the viscous timescale depends on local material properties. 
Translation of the viscous timescale of the event to the deposit radius depends on the viscosity of the material, and this, in turn, strongly depends on the disc temperature. The analysis of the spectra by \citet{petrushevska2023} did not lead to clear self-consistent results, so we analyze the broad-band photometric SED, which extends more toward UV.

Our toy model fits the profile best if multiple injections take place. It is thus likely that the star was not disrupted completely in one passage; instead, it experienced repeated episodes of mass loss while approaching the central black hole.
The disruption region is likely filled with a standard disc, as implied by the Eddington ratio of the source before the outburst. Still, the disruption event must have altered the state of the accreting material, and we reanalyze the Swift data, intending to have some constraints that could help link the timescales with the corresponding radii.
\section{Swift SED}
\label{sect:SED}

In this work, we used 3 observations from the Neil Gehrels Swift Observatory \citep{Gehrels2004} (Swift/UVOT) in UV bands of UVOT ($uvm1$, $uvw2$, $uvm2$) and optical $U$, $B$, and $V$ bands of UVOT, taken on MJD = 57632, 57647, and 57774 (PI: Dong; Target ID: 34704). We extracted the photometry in the same way as in \cite{petrushevska2023}. We present those observations in Fig. \ref{fig:ps16-few-sed}. The shape of SED did not change
significantly through observations, which were taken before the peak, and around peaks (see Fig. \ref{fig:lc-event}).

\begin{figure}
    \centering
      \includegraphics[scale=0.45]{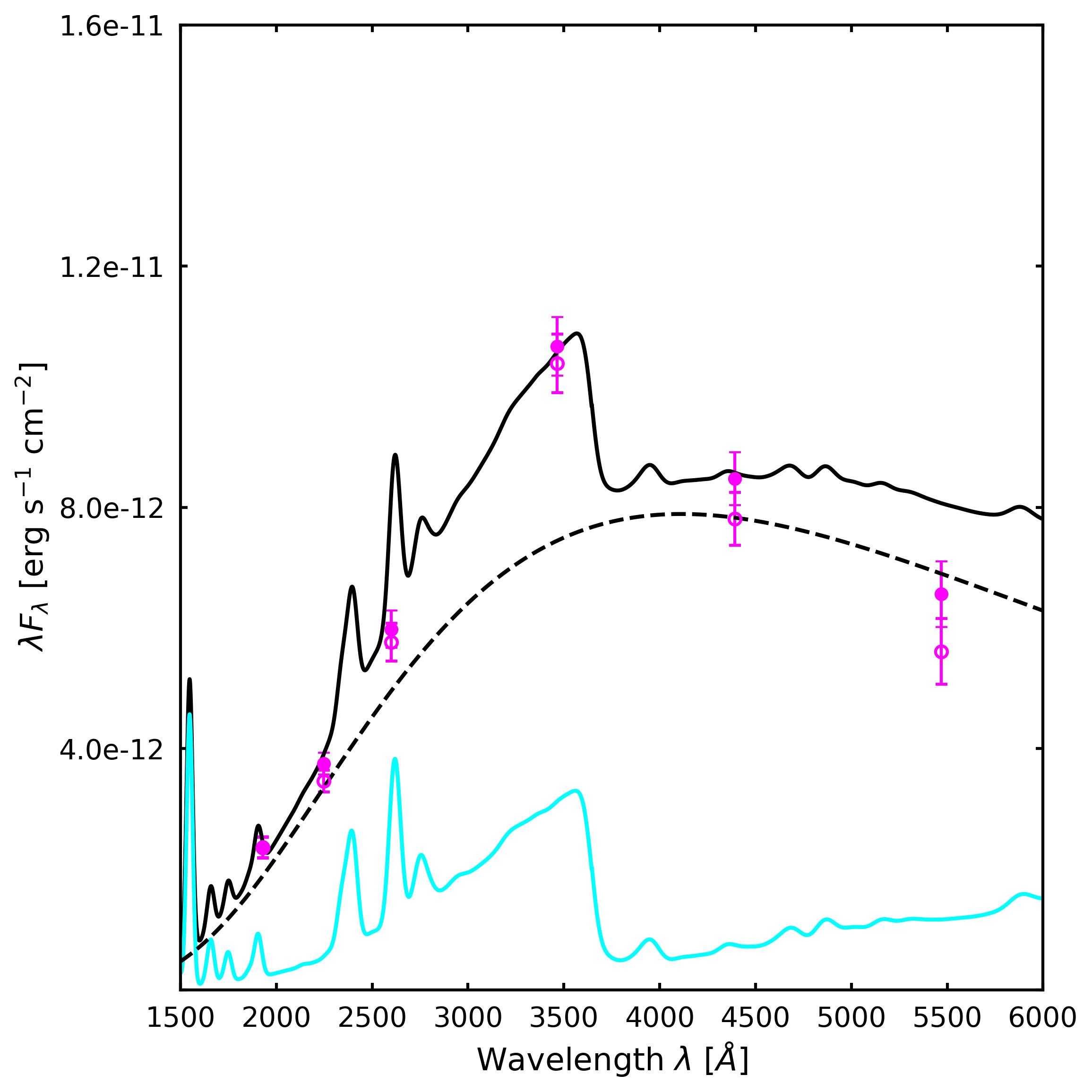}

    \caption{Rest-frame broad-band SED of the source from Swift (MJD = 57632) with (open circles) and without (filled circles) host subtraction. This spectrum is well fitted by a sum of a black body of the temperature 8,900 K (dashed line) and a contribution from the BLR (cyan). The Balmer edge well explains the excess at $\sim$ 3600 \AA.}
    \label{fig:swift-sed}
\end{figure}
We decided to model SED (and check SED shape with and without host subraction\footnote{For host subtraction, we used synthetic Host-Galaxy Magnitudes for \objname\ provided by \citet{2021hinkle}.}) taken before the MJD = 57632, and we present it in Figure~\ref{fig:swift-sed}. SED peaks in the near UV, and the shape does not correspond to the standard accretion disc. We tried to model it by assuming a \citet{ss73} accretion disc with absorption, but the fit was not satisfactory; the drop of the flux towards UV was not fast enough to represent the data. The same problem was encountered by \citet{petrushevska2023} when they analyzed the spectroscopic data. Our use of the Swift SED makes the situation clearer since the spectroscopic data starts at $\sim 3000$ \AA~ while we have the coverage starting from 1928 \AA.

\begin{figure*}
    \centering

\includegraphics[scale=0.15]{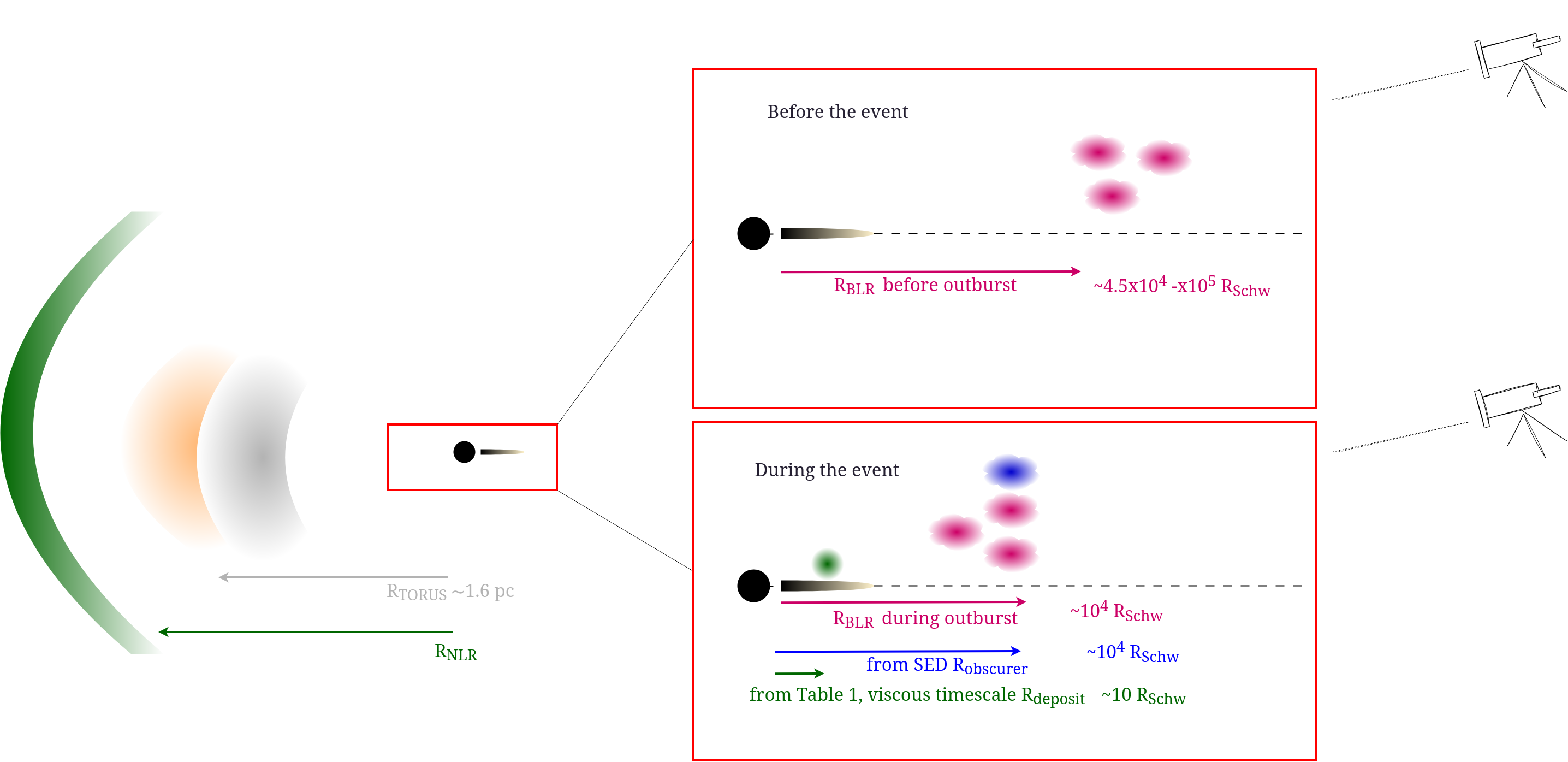}
\caption{Schematic view of the system. The central SMBH, accretion disc, BLR before the flare are shown in the upper zoom-in, and the BLR during the flare in the bottom zoom-in. In magenta, we mark the BLR, in blue, the obscurer through which we see the event. Using a green circle, we mark the location of the potential material deposits to the accretion disc. The orange and gray regions represent the torus and the evaporated part after the outburst of \objname\ \citep[see Fig. 3][]{2025jiang}, respectively. We mark the Narrow Line Region in green. The icons of the telescope indicate the viewing angle of the observer to the system. Note that the scale does not apply to this visualization. }
    \label{fig:schemat-postflare}
\end{figure*}

Surprisingly, we obtained a relatively good fit using a single black body shape. The requested temperature is very low, 8,900 K, not corresponding to a high accretion rate in the source. We see the peak of the SED very clearly, as well as the shape of the drop towards shorter wavelengths. To reconcile such a low temperature with the bolometric luminosity of the source, we must require a large emitting area, of the order of 7000 R$_{\rm{}Schw}$. The stellar disruption must have happened much closer in. We thus conclude that we do not actually see the disrupted region directly, and the observer is located behind some optically thick material produced during the disruption phase. The material, however, efficiently thermalizes and reprocesses the irradiation it must receive from the direct vicinity of a black hole. We illustrate it schematically in Figure~\ref{fig:schemat-postflare}. 

The SED shows a clear excess above the black body fit at $\sim 3600 $\AA. This is close to the Balmer edge. Since during the outburst, the BLR is clearly visible, we added this component to the fit. We used the shape of the characteristic BLR emission (Balmer continuum) using the CLOUDY software (version C23.00, \citealt{CLOUDY2023}), and we assumed parameters representative for single-zone BLR modelling: density $10^{11}$ cm$^{-3}$, and column density $3 \times 10^{23}$ cm$^{-2}$ \citep[see e.g.][and the references therein]{pandey2025}.
We normalized this spectral component arbitrarily, and with this component, the overall SED shape is well reproduced. \footnote{Modelling of full BLR emission profile including all emission lines is complex since clouds span a wide range of density and distance from the ionizing source \citep[see][]{1995ApJ...455L.119B, 2007ASPC..373..365L}.} The visibility of the BLR implies that only the central region is blocked from the observer, while the BLR is well visible, which we tried to illustrate in Figure~\ref{fig:schemat-postflare}.
We can roughly estimate the position of the emitter directly seen in the optical band. Assuming the bolometric luminosity $L_{\rm bol}$ of  10$^{44} \ergs$ \citep{blanchard2017}, and assuming the complete thermalization of the emission, we have a relation between the luminosity, temperature, and the radius of the obscurer: $L=4 \pi R_{\rm obsc}^2\sigma T^4$. The temperature of the reprocessor, from the SED fit, $T = 8900K$, implies  $R_{\rm obsc} = 4.7\times10^{15}\,{\rm cm}$, or 15800 in $R_{\rm{}Schw}$. 
Blocking the central source of radiation is not uncommon. Seyfert 2 galaxies with hidden BLRs represent one such class  \citep[e.g.][and the references therein]{2017pudu}, and among Galactic black hole binaries, SS 433 provides another example \citep{murdin1980}. This shielding of the central region implies that the temperature of the flow close to the black hole cannot be estimated, making it difficult to convert flow timescales into radii.
The fact of shielding the central parts implies that we have no estimate of the temperature of the flow close to the black hole, which makes the conversion of the flow timescales to radii difficult to constrain.

\section{Geometry of the event}
\label{sect:geom}

Since we cannot directly estimate the deposit radius for each of the mass components given in Table~\ref{tab:parameters} we perform only simple estimates of the consistency. Assuming the black hole mass from \cite{blanchard2017} we first calculate the orbital period $t_{orb}$ from the Keplerian velocity,
\begin{equation}
    \Omega_K(R_0) = \sqrt{GM_\mathrm{BH}/R_0^3}\;,
\end{equation}
where $t_{orb} = 2\pi/\Omega_K$ corresponding to 0.8h and 24h for representative values of the distance 10 $R_{\rm{}Schw}$ and 100 $R_{\rm{}Schw}$, respectively. Of course, the viscous timescales are much longer. 

Using $\alpha$-viscosity prescription for the kinematic viscosity $\nu$ we have the relation \citep{1998kato}
\begin{equation}
    \nu = 2/3 \alpha c_s^2 / \Omega_K(R_0)
\end{equation}
in which $\alpha$ is the viscosity parameter, $c_{\rm s}$ is the sound speed, and $\Omega_K(R_0)$ is the Keplerian angular velocity at radius $R_0$.
In the gas-pressure-dominated regime, the sound speed is directly related to the temperature, and we can adopt the hydrostatic equilibrium:

\begin{equation}
    c_s^2=\frac{kT}{m_p}.
\end{equation}
in which $c_s$ is the speed of sound, k is the Boltzmann constant, T is the temperature of the gas, and $m_p$ is the proton mass.
We use the viscous timescale (equation 6 from \cite{zdziarski2009}) as the decay timescale, and we substitute equations from above into $t_{visc}$
  \begin{equation}
   t_{visc}= \frac{2R_0^2}{3\nu}
      \label{eq:tvisc}
   \end{equation}
and we obtain
\begin{equation}
     t_{\rm visc} = \frac{\sqrt{GM_\mathrm{BH}R_0}}{\alpha \frac{kT}{m_p}}.
\end{equation}
From this equation, we finally have an expression for the deposit radius:
  \begin{equation}
     R_0 = \frac{t^2_{visc}\alpha^2{(\frac{kT}{m_p}})^2}{GM_\mathrm{BH}}.
\end{equation}

We assume $\alpha$ = 0.1, $M_\mathrm{BH} = 10^6\,M_{\odot}$, t$_{vis}$ = 1 yr, as representative values, and we estimate the deposit radius for a few potentially applicable values of the plasma temperature. The results are given in Table~\ref{tab:r0}. We see that the standard cold accretion disc model gives too long timescales, so the deposit would have to be inside the black hole horizon. We must consider much higher values of the temperature, close to $10^8$ K, for the plasma to explain its efficient flow towards the black hole. 
The disruption usually occurs at a few to a few tens of Schwarzschild radii. The final estimate is strongly dependent on the adopted viscosity. If the medium is highly turbulent, the viscosity coefficient $\alpha$ may be even higher than 0.1, shortening the viscous timescale. Higher temperature, which may result from the violent disruption of the object, can shorten the viscous timescale much more, pushing the disposal radius much more outward, although high exemplary values of the temperature in Table~\ref{tab:r0} are not likely. 

The firm outer limit for the initial event location is set by the obscurer through which the event is observed.
The position of the obscurer,  15800 $R_{\rm{}Schw}$, determined in Section~\ref{sect:SED}, thus sets the outer radius, and it is far beyond expectations for the inflow pattern. Thus, the position of the obscurer clearly implies outflow and does not provide any clue to the radial location of the disruption event. 

In addition, the overall structure of the source includes one more medium, even more distant than the obscurer.  \cite{2025jiang} reported an extreme giant dust structure with an inner radius of 1.6 pc, and its luminosity has been slowly rising after the outburst. This light echo appeared $\sim 200$ days after the event and continued to rise for a few years, with the peak luminosity estimated to be two orders of magnitude higher than the measured optical flux. However, this structure lies much farther out than the obscurer required to explain the Swift spectrum. It is instead consistent with the standard torus present around the AGN before the TDE, which is now more strongly irradiated due to the outburst. This opens a question about its nature.

To complement the location of the various structure elements of the source, we also put into Table~\ref{tab:r0} the position of the BLR radius. Its location before the outburst has been determined in Section~\ref{tab:sdss-measurements}.  The BLR radius during the outburst is, however, different. We can estimate it by taking the FWHM of the \hb~ as 5000 \kms \citep[see Fig. 17 in ][]{petrushevska2023}, and we obtain the new BLR distance 1.7 $\times 10^4 R_{\rm{}Schw}$. This is interesting: the value of the distance is of the same order as that of the obscurer. Clearly, the usual BLR cannot be so optically thick as to reprocess all the radiation from the inner region. However, part of the same medium, illuminated by the central source, can explain the birth of a new BLR. 
Thus, intending to comment more on the nature of the obscurer, we compare the mass of the obscurer to the mass of the event and to the mass of the inner accretion disc. We can estimate the mass of the obscurer assuming it must block and reemit radiation, so its optical depth $\tau_{\rm opt} \gtrsim 1$,

\begin{equation}
\tau_{\rm opt} = \kappa \Sigma,
\end{equation}
where $\Sigma $ is the surface density, and $\kappa$ is the opacity. For electron scattering $\kappa = 0.4$ cm$^2$g$^{-1}$, for dust and partially ionized medium, it will be higher. For simplicity, we can assume $\kappa = 1$ cm$^2$g$^{-1}$. If we set the distance of the obscurer at $R_{\rm obsc}$ , and introduce a covering factor f$_{\rm obsc}$, we have the expression for the mass of the obscurer, M$_{\rm obsc}$:
\begin{equation}
M_{\rm obsc} = f_{\rm obsc} 4 \pi R_{\rm obsc}^2 \Sigma,
\end{equation}
Taking the value of the absorber position from Table \ref{tab:r0}, and assuming $\tau_{\rm opt}$ = 1, $f_{\rm obsc} = 0.5$ we obtain M$_{\rm obsc}$ = 0.069\Msun, which is about 5 times smaller than disrupted mass (see Table \ref{tab:parameters}).

\begin{figure}
    \centering

              \includegraphics[scale=0.4]{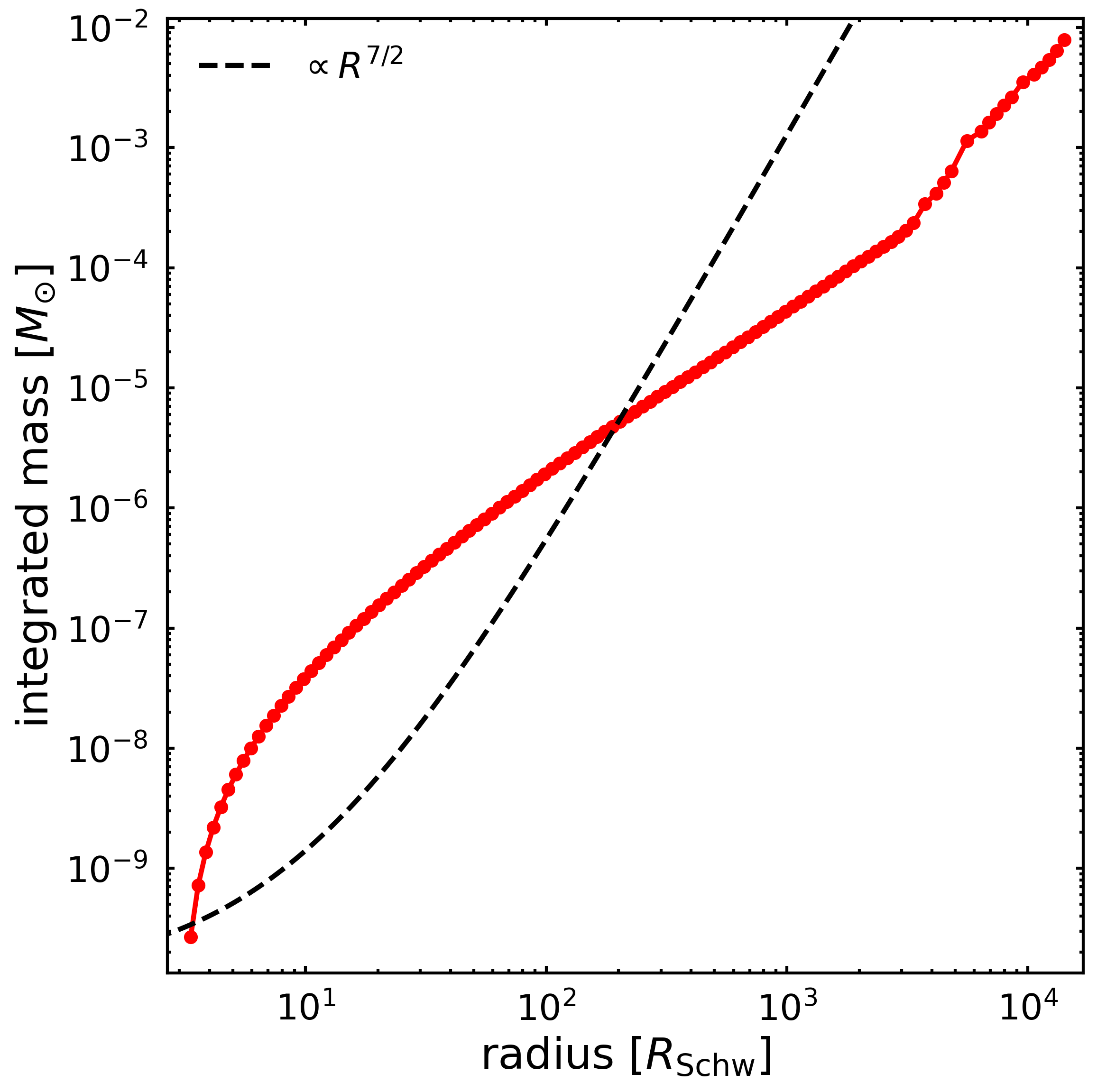}

    \caption{Integrated mass of the disc before the TDE as a function of the current radius, using $\alpha = 0.02$. To guide the eye, we also plot a power-law $\propto R^{7/2}$ expected for the region A solution in \cite{ss73} (the accretion disc is radiation-pressure dominated and the opacity is dominated by electron scattering). }
    \label{fig:disk_mass}
\end{figure}

In order to estimate the disc mass before the TDE, we used the standard accretion disc model, which includes all the necessary elements to calculate the disc vertical structure: complex opacity table, and both gas and radiation pressure \citep{rozanska1999}. The specific code version has been used in \citet{czerny2016}, but here we neglected the effects of self-gravity, which is unimportant in the case of a relatively low black hole mass. We also did not include any effects of the magnetic field. We calculated the disc surface density for each radius, and integrated it from the inner radius (see Figure~\ref{fig:disk_mass}). 
The total disc mass integrated up to the distance of the obscurer is 0.079 $M_{\odot}$, actually comparable to the TDE mass. 
Both the mass of the disc before the disruption event and the disrupted mass are comparable, within a factor of a few, to the mass of the obscurer.
Therefore, we cannot argue whether the obscurer was formed from the disrupted material or from the disc material pushed out by the sudden radiation pressure created by the disruption. In both cases, the material from the innermost region must reach the distance of the obscurer on the timescale of a few months. This requires the velocity of the order of a few thousand km s$^{-1}$, which can likely be achieved in such an event.

\begin{table}
\centering
\caption{Conversion of the viscous timescale of 1 year in the disc around  \objname\ for exemplary values of the plasma temperature, expected from a cold disc, warm corona or hot corona. }
\label{tab:r0}
\begin{tabular}{llll}
\hline
\hline
Zone&Temperature &  R$_0$    &  R$_0$    \\ 
 & [K] & [R$_{\rm{}Schw}$] & [pc]  \\
\hline
Cold disc       &  10$^5$      &  $<<$ 1  & -  \\
Warm corona  &  10$^6$     & 0.0025  & - \\

Warm corona  &  10$^7$     & 0.25  & - \\
Warm corona  &  10$^8$     & 25  & -\\
Hot corona  &  $10^9$     & 2500 &    \\
Obscurer  &  -  & 15800 & $1.52 \times 10^{-3}$\\
BLR during event &  -  &  $10^4$ & $9.64\times 10^{-4}$\\
BLR before event & - &  $10^5$ & $9.64 \times 10^{-3}$\\
Dusty structure & -   &  $1.7 \times 10^7$  & 1.6  \\
\hline     
\end{tabular}
\end{table}

%


\begin{table*}
\centering
\caption{Estimates of the tidal disruption radius around a $10^6 M_{\odot}$ black hole for several types of astronomical bodies.}
\label{tab:disruption-radius-astronomical-bodies}
\begin{tabular}{lccc}
\hline
\hline
Object& Radius [km]& Mass $[M_{\odot}$] & disruption radius   [R$_{\rm{}Schw}^*$] \\ \hline
Sun & 6.96$\times 10^5 $  &  1 & 23.6\\
M-type dwarf    &$ 2.7 \times 10^5 $   &  0.3 & 13.7\\
white dwarf  &$7\times 10^3$ & 0.6 & 0.28\\
neutron star &10& 1.5& 3$\times 10^{-4}$\\
core of red giant star &7$\times 10^3$ &0.4&  0.32\\
shell of red giant star &5$\times 10^5$ &0.3&  25.3\\

brown dwarf & $1.75 \times 10^4$   &   0.01 & 2.8\\
\hline     
\end{tabular}
\begin{center}
\footnotesize{*R$_{\rm{}Schw}$ of central black hole.}
\end{center}
\end{table*}




\section{The nature of the disrupted body}
\label{sec:nature-of-dusruption}
In this section, we explore possible types of astronomical bodies that could have been disrupted and processes in star-AGN accretion disc interaction.
\subsection{Possible identities of the disrupted body}
Since we have no firm constraints on the radial location of the event from previous considerations, we evaluate the possible nature of the event by considering the stars that could have been disrupted.  For the star with the radius $R_{\star}$ and the mass of $m_{\star}$, the tidal disruption radius $r_{\rm t}$ is $r_{\rm t}\sim R_{\star}(M_\mathrm{BH}/m_{\star})^{1/3}$, which in Schwarzschild radii can be expressed as
\begin{align}
    \frac{r_{\rm t}}{R_{\rm Schw}} &\sim \frac{c^2}{2G} R_{\star} m_{\star}^{-1/3} M_\mathrm{BH}^{-2/3}\,\notag\\
    &\sim 23.6 \left(\frac{R_{\star}}{1\,R_{\odot}} \right) \left( \frac{m_{\star}}{1\,M_{\odot}} \right)^{-1/3} \left(\frac{M_\mathrm{BH}}{10^6\,M_{\odot}} \right)^{-2/3}.
    \label{eq_tidal_radius1}
\end{align}
 For main-sequence stars, including M-type stars, we have $R_{\star}\propto m_{\star}^{0.8}$ \citep{2017imas.book.....C}, which leads to the numerical estimate for $r_{\rm t}$,
\begin{equation}
   r_{\rm t} \sim 13.45 \left( \frac{m_{\star}}{0.3\,M_{\odot}} \right)^{7/15} \left(\frac{M_\mathrm{BH}}{10^6\,M_{\odot}} \right)^{-2/3}  R_{\rm Schw}\,,
   \label{eq_tidal_radius_ms}
\end{equation}
 which is scaled to $m_{\star}=0.3\,M_{\odot}$ in agreement with the value inferred from the TDE light-curve fitting. Hence, the lighter the main-sequence star is, the closer its tidal radius is with respect to the SMBH.
In Table \ref{tab:disruption-radius-astronomical-bodies}, we compare estimates of the tidal disruption radius for different astronomical bodies. Due to extreme density contrast between the core and the shell of giant stars, we consider the core and the shell separately. Compact objects like neutron stars, white dwarfs, or brown dwarfs are not plausible candidates for the event, as they are not disrupted by the central body. Similarly, the dense core of a red giant is unlikely to be disrupted. Most interesting results (with disruption radii of $\sim 13.7\,R_{\rm{}Schw}$ and $25.3\,R_{\rm Schw}$) were obtained for the M-type dwarf and the shell of the red giant star.

Giant stars have a much lower average density 
and can be disrupted and lose part of the envelope at larger distances than main-sequence stars \citep[][and references therein]{navarro2025}. It is challenging to estimate the mass contained in the envelope without specific numerical simulations. Simulations performed by \citet{navarro2025} imply that for the disruption radius $\sim 30 R_{\rm{}Schw}$ the stripped mass would be from $\sim 0.3 M_{\odot}$ for a $1 M_{\odot}$ star to $\sim 1 M_{\odot}$ for more massive, $2 M_{\odot}$ star. Thus, a lower-mass red giant seems to be more consistent with the observational constraints for the object \objname.
However, the Asymptotic Giant Branch (AGB) phase is shorter in comparison to the Main Sequence (MS) phase, making TDEs involving AGB stars less likely. Alternatively, a low-mass main-sequence star with a mass of $\sim 0.3 M_{\odot}$ would be fully disrupted, leaving no surviving remnant. Since such low-mass stars are more numerous than other stellar types, they are the most likely candidates for disruption.


\subsection{Timescales, dynamics and geometrical constraints}
\label{subsec_timescales_dynamics}

\begin{figure}
    \centering
    \includegraphics[width=\columnwidth]{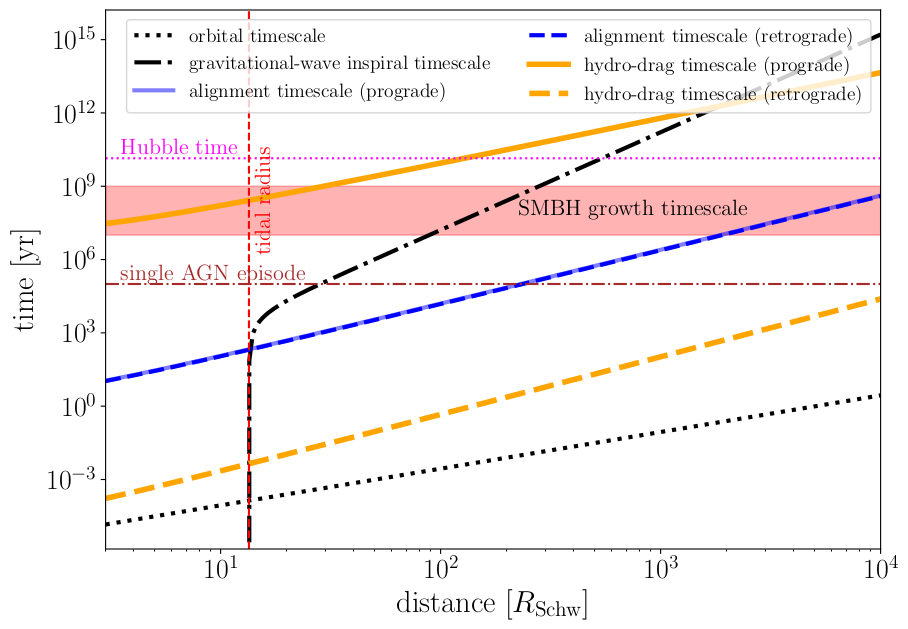}
    \caption{Alignment, hydrodynamical drag, gravitational-wave inspiral, and orbital timescales (in years) as a function of the distance from the SMBH (in Schwarzschild radii) for a $0.3\,M_{\odot}$ star. For the accretion-disc properties, we adopt the relative accretion rate of $\dot{m}=0.1$. In addition, we depict the typical AGN lifetime ($\sim 10^7-10^9$ years, shaded red region), a single AGN episode ($\sim 10^5$ years; dash-dotted brown line), the Hubble time (horizontal dotted magenta line), and the tidal radius (vertical dashed red line).}
    \label{fig_timescales}
\end{figure}

In this subsection, we examine dynamical processes that drive the inspiral of stars in AGN accretion discs and associated timescales.
When an inclined star of mass $m_{\star}$ interacts with an accretion disc, it is subject to hydrodynamical drag as it crashes with the accretion disc twice along its orbit. Due to the angular momentum change due to the disc drag, the star is expected to align with the disc plane on the so-called alignment (grinding) timescale \citep{1991MNRAS.250..505S},
\begin{align}
   t_{\rm align} (\iota_0 \rightarrow \iota_1) &\sim \frac{m_{\star}P_{\rm orb}}{\pi R_{\star}^2 \Sigma_{\rm disc}}\times\,\notag\\
   &\times\left[\ln{\left(\frac{1+\sqrt{1-\sin^2{(\iota/2)}}}{\sin{(\iota/2)}} \right)} -2\sqrt{1-\sin^2{(\iota/2)}}\right]_{\iota_0}^{\iota_1}\,, 
   \label{eq_alignment_timescale}
\end{align}
where $\pi R_{\star}^2$ is the cross-section of the star, $\Sigma_{\rm disc}$ is the surface-density of the standard accretion disc, which we consider in this and the following estimates. In addition, $P_{\rm orb}$ is the orbital period of the star, assuming a bound, elliptical orbit. For $t_{\rm align}$, we consider stars that are nearly corotating, inclined initially at $\iota_0=45^{\circ}$, and the final state is evaluated at the grazing orbit with $\iota_{1}\sim H_{\rm disc}/r$ where $H_{\rm disc}$ is the disc scale-height and $r$ is the distance of the star from the SMBH. For $r\in (3,10\,000)\,R_{\rm Schw}$, we obtain $t_{\rm align}\sim 11.1-4.1\times 10^8$ years, see Fig.~\ref{fig_timescales}. For nearly retrograde stellar orbits, the alignment timescale from $\iota_{0}=\pi-H_{\rm disc}/r$ to $\iota_1=H_{\rm disc}/r$ is a bit smaller due to the much larger initial relative velocity and hence the drag; in the same distance range, we obtain  $t_{\rm align}\sim 10.7-3.9\times 10^8$ years.

\begin{figure}
    \centering
    \includegraphics[width=\columnwidth]{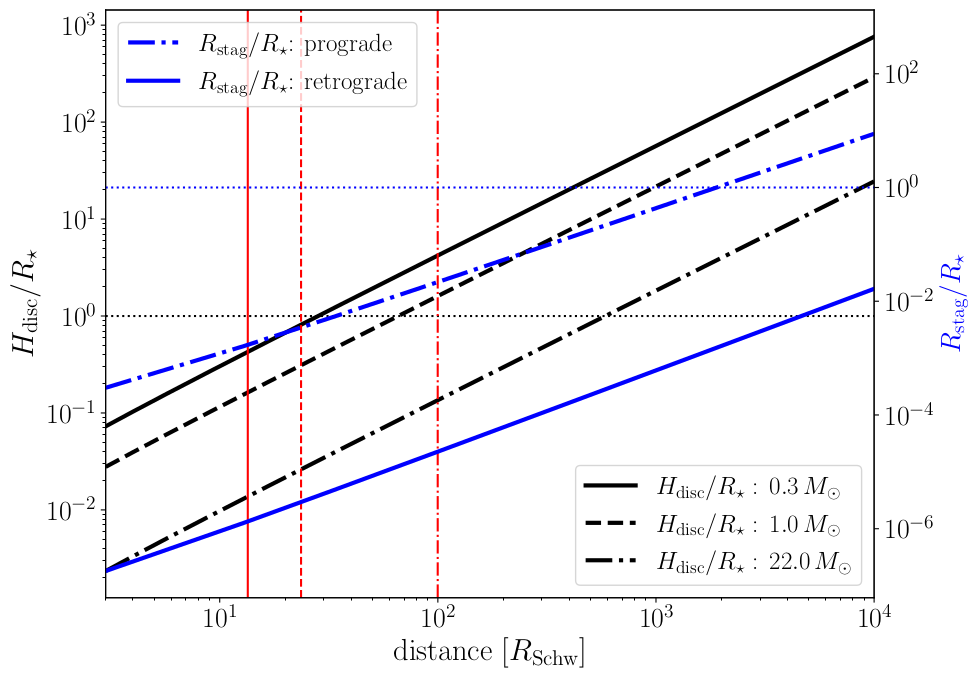}
    \caption{Standard accretion disc scale-height to stellar radius ratio as a function of distance from the SMBH (in Schwarzschild radii) for main-sequence stars with $m_{\star}=0.3\,M_{\odot}$ (solid) and $m_{\star}=1\,M_{\odot}$ (dashed) -- left axis. When the star is disrupted at the corresponding tidal radius (highlighted by solid and dashed red vertical lines), its radius is thicker than the disc scale-height. In this figure, we also show the ratio of the stagnation radius of the stellar bow-shock to the effective stellar radius, $R_{\rm stag}/R_{\star}$ -- right axis. For numerical estimates we adopted the massive star of $10\,M_{\odot}$ with the extreme mass-loss rate of $\dot{m}_{\rm w}\sim 10^{-5}\,{\rm M_{\odot}\,yr^{-1}}$ and the terminal wind velocity of $780\,{\rm km\,s^{-1}}$. $R_{\rm stag}/R_{\star}$ is depicted for the coorbiting (prograde) case (blue dash-dotted line) as well as for the counter-orbiting (retrograde) case (blue solid line). Black and blue dotted horizontal lines stand for unity values for the corresponding quantities ($H_{\rm disc}/R_{\star}$ and $R_{\rm stag}/R_{\star}$) on the left and the right axes, respectively.}
    \label{fig_HR}
\end{figure}

Once the star is aligned with the accretion disc, it is subject to hydrodynamical drag force due to the surrounding disc gas, which depends on the relative velocity of the star with respect to the disc gas, $v_{\rm rel}=[2GM_\mathrm{BH}/r(1-\cos{\iota})]^{1/2}$, where $\iota$ stands for the inclination of the star with respect to the disc plane ($\iota=0^{\circ}$ for the co-rotating star and $\iota=180^{\circ}$ for the counter-rotating star). The stellar body of a $\sim 0.3\,M_{\odot}$ main-sequence star is expected to interact directly with the disc gas since the minimum stellar-wind velocity required to counter-balance the ram pressure,
\begin{equation}
    v_{\rm w,lim}\leq \frac{4\pi R_{\star}^2 \rho_{\rm disc} c_{\rm s}^2}{\dot{m}_{\rm w}}\,,
    \label{eq_wind_limit}
\end{equation}
is of the order of $v_{\rm w,lim}\sim 10^{15}\,{\rm km\,s^{-1}}$ at $r_{\rm t}\sim 13.45\,{\rm R_{\rm Schw}}$ (see Eq.~\eqref{eq_tidal_radius_ms}), which exceeds by twelve orders of magnitude the expected terminal wind velocity of the M-dwarf star ($v_{\rm w}\sim 550\,{\rm km\,s^{-1}}$; here we assumed that the star is comoving within the standard Shakura-Sunyaev disc solution with the relative accretion rate of $\dot{m}=0.1$; from this we calculated the disc mass density $\rho_{\rm disc}$ and the disc-gas sound speed $c_{\rm s}$; for the M-dwarf star, the expected mass-loss rate $\dot{m}_{\rm w}\sim 10^{-14}\,{\rm M_{\odot}\,yr^{-1}}$ is comparable to the Solar value). Even if the star accretes mass and reaches $m_{\star}\sim 10\,M_{\odot}$ (OB spectral type) with the extreme mass-loss rate of $\dot{m}_{\rm w}=10^{-5}\,M_{\odot}\,{\rm yr^{-1}}$ and the terminal (escape) wind velocity of $780\,{\rm km\,s^{-1}}$, the stagnation radius would not exceed the stellar radius up to $r\sim 1900\,R_{\rm Schw}$ for the coorbiting star, see Fig.~\ref{fig_HR} (blue dot-dashed line, right axis). For the counter-orbiting star with the same parameters (blue solid line, right axis), the stagnation radius is smaller than the expected stellar radius up to $r\sim 174\,000\,R_{\rm Schw}$, hence even the massive counter-orbiting stars will be ablated from their photospheres across the whole radial range of the accretion disc. 

The star is, however, not fully embedded in the thin accretion disc since the disc scale-height/stellar radius ratio is $H_{\rm disc}/R_{\star}\sim 0.43$ at $r_{\rm t}\sim 13.45\,{\rm R_{\rm Schw}}$ for the standard disc solution (the star is about 2.3-times thicker than the disc). The ratio is $H_{\rm disc}/R_{\star}\sim 1$ at $r\sim 28.2\,R_{\rm Schw}$ for the relative accretion rate of $\dot{m}\sim 0.1$. We plot the dependency of $H_{\rm disc}/R_{\star}$ on the distance from the SMBH for $m_{\star}=0.3\,M_{\odot}$ and $m_{\star}=1\,M_{\odot}$ main-sequence stars in Fig.~\ref{fig_HR}, where we assume that $R_{\star}\sim m_{\star}^{0.8}$ for simplicity. Since the disrupted star is typically geometrically thicker at the corresponding tidal radius, it has implications for the subsequent evolution of the stellar debris since they are freer to expand above and below the disc plane than within the disc.

Since the stellar wind is too weak to counter-balance the ambient-medium ram pressure, we can use the hydrodynamic-drag relation for an inert spherical object of radius $R_{\star}$. For the hydrodynamic drag acting on the star within the disc plane, we adopt the following relation for the drag force $F_{\rm drag}\simeq -\pi R_{\star}^2 \rho_{\rm disc}v_{\rm rel}^2$ \citep{1999ApJ...513..252O,2000ApJ...536..663N}, where the negative sign expresses the fact that the drag force has an opposite direction with respect to the relative velocity of the star. The star will thus lose its orbital energy due to the drag on the $e$-folding timescale that can be estimated as \citep{2000ApJ...536..663N},
\begin{equation}
    t_{\rm drag}\simeq \frac{E}{|\dot{E}|}\sim \frac{m_{\star} v_{\star}}{\pi R_{\star}^2 \rho_{\rm disc} v_{\rm rel}^2}\,.
\end{equation}
This implies that the hydrodrag is maximized ($t_{\rm drag}$ is the smallest) for the counter-rotating case when $v_{\rm rel}\sim 2v_{\star}$, where $v_{\star}$ stands for the local orbital velocity. The hydrodrag timescale also gets smaller for a larger disc density (larger accretion rate) and decreases with the stellar mass as $t_{\rm drag}\propto m_{\star}^{-3/5}$ for main-sequence stars. On the other hand, $t_{\rm drag}$ increases significantly when the star comoves with the disc and the relative velocity with respect to the disc motion, $v_{\rm rel}=|v_{\star}-v_{\rm disc}|\sim \delta v_{\rm star}$ where $\delta\sim (H_{\rm disc}/r)^2$. In Fig.~\ref{fig_timescales} we compare $t_{\rm drag}$ for the prograde case (solid orange line) and the retrograde case (dashed orange line) -- they differ by nine to eleven orders of magnitude from larger to smaller distances.

We compare the alignment and hydrodynamical-drag timescales with the timescale related to the gravitational-wave inspiral from an initial radius $r_0$ to the tidal disruption radius $r_{\rm t}$ \citep{1964PhRv..136.1224P},
\begin{equation}
    t_{\rm GW}\simeq \frac{5c^5}{256 G^3} \frac{(r_0^4-r_{\rm t}^4)}{M_\mathrm{BH}m_{\star}(M_\mathrm{BH}+m_{\star})}\,,
    \label{eq_gw_timescale}
\end{equation}
where $r_{\rm t}$ is approximately given by Eqs.~\eqref{eq_tidal_radius1} and \eqref{eq_tidal_radius_ms}. The expression~\eqref{eq_gw_timescale} assumes a nearly circular orbit.

In Fig.~\ref{fig_timescales} we compare all the relevant timescales as a function of distance from the SMBH (in Schwarzschild radii) that are related to the dynamical processes affecting the stellar orbit close to the actively accreting SMBH. For comparison, we also depict the typical timescale for a single AGN episode \citep[$\sim 10^5$ years in comparison with the total SMBH growth timescale of $\sim 10^7-10^9$ years,][]{2015MNRAS.451.2517S}, the Hubble time, and the tidal radius of a $0.3\,M_{\odot}$ star. For the accretion disc properties (density, temperature, and scale-height) we adopt the standard disc solution with the relative accretion rate of $\dot{m}=0.1$. It is apparent that during the AGN lifetime of $\sim 10^5$ years, any misaligned stellar orbit with the orbital radius of $\lesssim 232-237\,R_{\rm Schw}$ can be brought into the disc plane via the star-disc interaction. Hence, these stars are predominantly expected to be aligned and corotating with the disc. However, for aligned corotating stars, the hydrodrag is not efficient due to a small relative velocity, hence the main inspiral mechanism is the rather slow gravitational-wave emission, which can effectively bring in low-mass stars towards the tidal radius that have initial orbital radii smaller than $r\lesssim 542\,R_{\rm Schw}$. Otherwise, the inspiral would take longer than the Hubble time. A faster inspiral of comoving stars can be provided by density waves within the accretion disc \citep{2004MNRAS.354.1177S}, however, we do not consider them in the current analytical estimates. A counter-rotating star, if it is embedded by chance within the disc from the onset of an AGN activity, can inspiral towards the tidal radius rather fast, $t_{\rm drag}\lesssim 25\,000$ years within $r\sim 10\,000\,R_{\rm Schw}$.   

\begin{figure}
    \centering
    \includegraphics[width=\columnwidth]{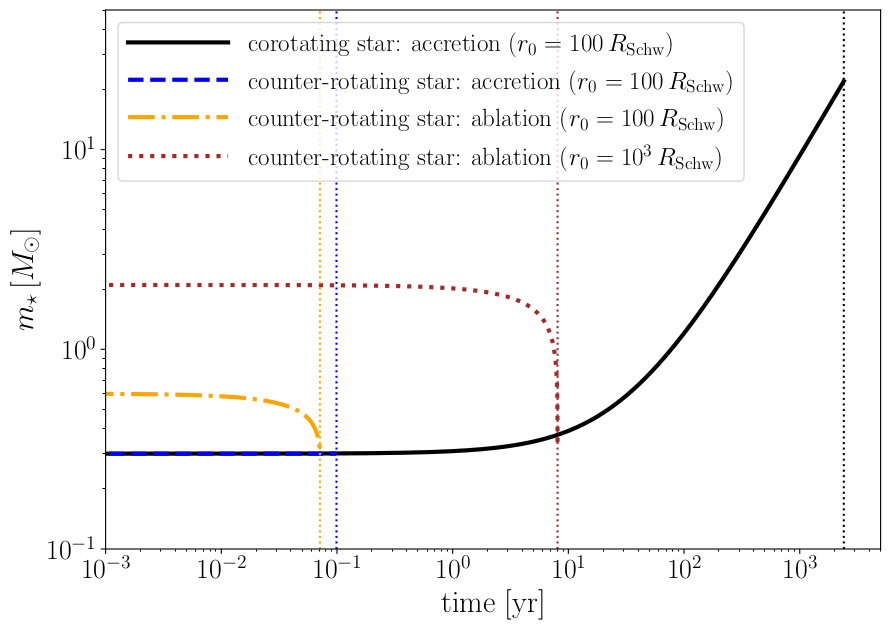}
    \caption{Stellar mass temporal evolution for a star embedded within the accretion disc for the corotating case when it mostly accretes mass (black solid line; initial mass of $0.3\,M_{\odot}$ at $100\,R_{\rm Schw}$ from the SMBH). We also depict the counter-rotating case without ablation (blue dashed line; initial mass of $0.3\,M_{\odot}$ at $100\,R_{\rm Schw}$ from the SMBH), the counter-rotating case with accretion-disc ablation included (orange dash-dotted line; initial mass of $0.6\,M_{\odot}$ at $100\,R_{\rm Schw}$), and the counter-rotating case with ablation that starts with the larger stellar mass at the larger distance (brown dotted line; initial mass of $2.1\,M_{\odot}$ at $1000\,R_{\rm Schw}$). The calculation involves the inspiral due to the gravitational-wave emission and the hydrodynamic drag due to the accretion disc and stops at the moment when the star reaches its corresponding tidal radius (depicted by vertical dotted lines), which increases rapidly for the corotating star whose mass and radius get larger due to rapid accretion. The relative accretion rate is set to $\dot{m}=0.1$.}
    \label{fig_stellar_mass}
\end{figure}

The orientation of the aligned orbit can be argued to be more likely counter-rotating with respect to the accretion disc. Although this is not expected for the long-term AGN activity since the stars should align with the disc and corotate, the star can be counter-rotating and embedded within the disc by chance due to an isotropy of the Nuclear Star Cluster. For such a retrograde orbit, the hydrodynamic drag is more efficient by ten orders of magnitude and has an $e$-folding timescale of the order of $\lesssim 25\,000$ years (retrograde case, see Fig.~\ref{fig_timescales}) for a large range of orbital radii up to $r\sim 10^4\,R_{\rm Schw}$, where stars can be located within the Nuclear Star Cluster \citep[e.g. the Milky Way case around Sgr A*;][]{2009ApJ...692.1075G,2020ApJ...899...50P} and the disc can also extend to these length-scales. Hence, even during a single AGN episode, stars counter-orbiting with respect to the accretion disc, can inspiral towards the tidal radius and get disrupted. For a prograde case, the stellar orbit of a low-mass star would need to be initially at $\lesssim 500\,R_{\rm Schw}$ for the gravitational-wave inspiral to take place within the Hubble time, which is more dynamically restricted. Since the alignment timescale is always smaller than the gravitational-wave inspiral timescale, the star can align relatively fast with different accretion disc orientations. For the last AGN episode, a low-mass star if comoving with the disc after alignment would have to be initially located at $\lesssim 30\,R_{\rm Schw}$ so that gravitational-wave inspiral could bring it to the corresponding tidal radius within the duration of the AGN activity.

The second argument is based on the inferred mass of the disrupted star of $\sim 0.3\,M_{\odot}$. If such a star were comoving with the disc, e.g. at the initial distance of $r\sim 100\,R_{\rm Schw}$ and with the initial mass of $\sim 0.3\,M_{\odot}$, it would gain mass during the inspiral via gravitational capture of the surrounding gas from the accretion disc. The mass-accretion rate is given by $\dot{M}_{\rm acc}\simeq \pi R_{\rm acc}^2 (c_{\rm s}^2 + v_{\rm rel}^2)^{1/2} \rho_{\rm disc}$, where the influence radius $R_{\rm acc}$ is the minimal value of the gravitational capture (Bondi) radius and the Hill radius, $R_{\rm acc}=\text{min}[R_{\rm cap},R_{\rm Hill}]$ \citep{1991MNRAS.250..505S}. The gravitational capture radius is given by $R_{\rm cap}=2 G m_{\star}/(v_{\rm rel}^2+c_{\rm s}^2)$ and the Hill radius can be expressed as $R_{\rm Hill}=r[m_{\star}/(3M_\mathrm{BH})]^{1/3}$. In our calculations, we include the standard accretion disc with the fixed relative accretion rate of $\dot{m}=0.1$. In addition, we include the gravitational-wave emission term \citep{1964PhRv..136.1224P},
  \begin{equation}
    \frac{\mathrm{d}r}{\mathrm{d}t}=-\frac{64 G^3 M_\mathrm{BH}m_{\star}(M_\mathrm{BH}+m_{\star})}{5c^5 r^3}\,,
    \label{eq_gw_term}
  \end{equation}
as well as the hydrodynamic-drag term,
  \begin{equation}
    \frac{\mathrm{d}r}{\mathrm{d}t}=-\frac{4 \pi r^{1/2}(GM_\mathrm{BH})^{1/2}R_{\star}^2 \rho_{\rm disc}(1-\cos{\iota})}{m_{\star}}\,.
    \label{eq_hydrodrag_term}
  \end{equation}
At every timestep, we trace the mass of the star affected by accretion, its radius, which both influence the tidal radius where it gets disrupted. We stop the integration when the star reaches its corresponding tidal radius.

In Fig.~\ref{fig_stellar_mass}, we plot the mass evolution for the coorbiting star during the gravitational-wave inspiral from the initial distance of $r=100\,R_{\rm Schw}$ that lasts for $\sim 2400$ years until the star reaches its tidal radius (black solid line). During that time, the mass of the star would eventually be $m_{\star}\sim 22\,M_{\odot}$ and therefore its tidal radius would be positioned at $r_{\rm t}\sim  99.86\,R_{\rm Schw}$ (assuming the main-sequence radius-mass relation). This is in contradiction with the inferred mass for the accretion event \objname, which is well below one Solar mass. In comparison, for the counter-orbiting star, the mass remains rather unchanged due to limited accretion (blue dashed line; no ablation is included in this case). The star would inspiral rather quickly within $\sim 0.1$ years from $r=100\,R_{\rm Schw}$ to its tidal radius at $\sim 13.45\,R_{\rm Schw}$ due to the strong hydrodynamical drag within the accretion disc.

The effect of ablation due to the accretion disc is especially relevant for the counter-orbiting case since the stellar wind cannot typically counteract the ram pressure of the ambient medium and thus create a gap along the orbit, see Eq.~\eqref{eq_wind_limit}. The mass-loss rate due to ablation can be inferred from the balance of the incoming and outgoing momentum fluxes. The incoming momentum flux is related to the swept-up disc mass, $\dot{p}_{\rm in}\sim \dot{m}_{\rm sweep}v_{\rm rel}\sim \rho_{\rm disc}\kappa \pi R_{\star}^2 v_{\rm rel}^2$, where $\kappa\sim H/R_{\star}$ is the covering factor of the star by the disc, see Fig.~\ref{fig_HR}, which is less than unity close to the tidal radius. The outgoing momentum flux can be inferred from the rate of ablated mass that moves close to the escape speed from the star, $\dot{p}_{\rm out}\sim \dot{m}_{\rm abl} v_{\rm esc}$, where $v_{\rm esc}=(2 G m_{\star}/R_{\star})^{1/2}$. Only the fraction $\epsilon$ of the incoming momentum flux is effectively used to remove the mass. Then the mass-ablation rate can be expressed as 
\begin{equation}
    \dot{m}_{\rm abl}=\epsilon \kappa \rho_{\rm disc} \pi R_{\star}^2 \frac{v_{\rm rel}^2}{v_{\rm esc}}\,,
\end{equation}
which is close to zero for corotating cases ($v_{\rm rel}\sim 0$). In the calculations, we set $\epsilon=0.01$, which leads to the initial smooth decrease in stellar mass at larger distances from the SMBH. In Fig.~\ref{fig_stellar_mass} we show that when the ablation by the disc is included for counter-rotating cases, the mass of stars that have initially $0.6-2.1\,M_{\odot}$ at the distance of $100-1000\,R_{\rm Schw}$ is reduced rather quickly within $0.07-8$ years, respectively, to the value close to $0.3\,M_{\odot}$ (see orange dot-dashed and brown dotted lines). This is due to the maximized ram pressure for counter-orbiting stars and at the same time the star inspirals to denser regions of the disc due to the hydrodynamic drag. Hence, the inferred disrupted mass of $\sim 0.3\,M_{\odot}$ could plausibly be the remnant core of an initially more massive star that has undergone rapid ablation within the accretion disc.

\begin{figure*}
    \centering
    \includegraphics[width=\textwidth]{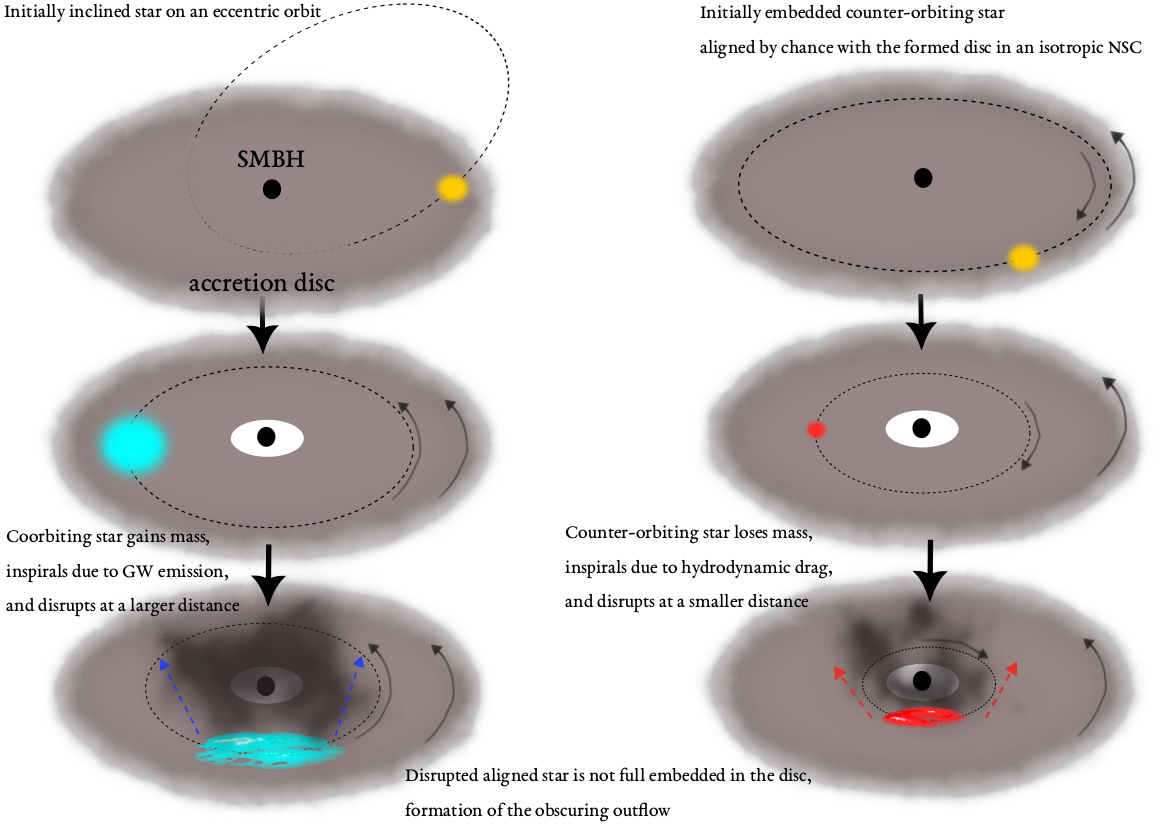}
    \caption{Illustration of the likely sequence of dynamic processes taking place in the star-AGN accretion disc interaction. At the top, the star is initially moving on an inclined eccentric orbit (left) or is already embedded within the disc on a retrograde orbit by chance (right). The orbit gets aligned and circularized due to the hydrodynamic drag. The fate of the star then depends on the sense of its orbit: if it is coorbiting with the disc (left), it gains mass via gravitational capture, inspirals due to the gravitational-wave emission, and finally gets disrupted at a larger distance. In contrast, for a counter-orbiting star (right), the star tends to lose mass rather than increase it due to fast ablation. It continues to inspiral swiftly due to hydrodynamical drag and eventually gets disrupted with a smaller mass and at a smaller distance in comparison with the coorbiting case. The latter, counter-orbiting scenario appears to be more consistent with the \objname\ event. }
    \label{fig_sketch_TDE}
\end{figure*}

In Fig.~\ref{fig_sketch_TDE}, we illustrate the dynamical process of alignment with the accretion disc and the two cases of the tidal disruption that depend on the sense of the stellar orbit: if the star is coorbiting (left), it gains mass via gravitational capture and is tidally disrupted at a larger distance; if, on the other hand, it is counter-orbiting and embedded within the accretion disc by chance from the onset of an AGN activity, it will tend to lose mass via ablation and it will quickly inspiral due to hydrodrag. In the end, the counter-orbiting star will get tidally disrupted at a significantly smaller distance than the co-orbiting star. 

Under the assumption of the circularization due to gas drag, the star orbits around the SMBH on an approximately circular orbit, presumably in an opposite sense to the accretion disc, which can consistently address the low mass of a disrupted star. At the time of the TDE the star is located at the distance from the SMBH corresponding to the tidal radius, $r_{\rm t}\sim R_{\star} (M_\mathrm{BH}/m_{\star})^{1/3}$, which can numerically be scaled to the Solar mass and expressed in Schwarzschild radii, $r_{\rm t}\sim 23.6\,R_{\rm Schw}$, see Eq.~\eqref{eq_tidal_radius1}.
Assuming the relation for lighter main-sequence stars, we have $R_{\star}\propto m_{\star}^{4/5}$, which gives the numerical estimate for $m_{\star}\sim 0.3\,M_{\star}$, $r_{\rm t}\sim 13.45\,R_{\rm Schw}$, see Eq.~\eqref{eq_tidal_radius_ms}.
At the distance of $r_{\rm t}$, the orbital timescale of a star or its debris is,
\begin{align}
    P_{\rm orb} &= \frac{2^{5/2}\pi GM_\mathrm{BH}}{c^3} \left(\frac{r_{\rm t}}{R_{\rm Schw}} \right)^{3/2}\,\notag\\
    &=1.2 \left(\frac{M_\mathrm{BH}}{10^6\,M_\mathrm{BH}} \right) \left(\frac{r_{\rm t}}{13.45\,R_{\rm Schw}} \right)^{3/2}\,\text{hours}\,.
\end{align}
Hence, if only one star is responsible for the TDE and there were at least two accretion events, they occurred within $\sim 0.01\,{\rm yr}/P_{\rm orb}\sim 73$ orbital timescales from each other (see deposit times in Table~\ref{tab:parameters}). Since the tidally disrupted star is stretched along its orbit by a factor of 2 to $2R_{\star}$ on the tidal timescale \citep{2012ApJ...755..155S,2022ApJ...931...39M},
\begin{align}
    \tau_{\rm tidal} &\approx \frac{r^{3/2}}{\sqrt{GM_\mathrm{BH}}}\sim 0.19\,\left(\frac{M_\mathrm{BH}}{10^6\,M_{\odot}} \right)\left(\frac{r}{13.45\,R_{\rm Schw}} \right)^{3/2}\text{hours}
\end{align}
which is comparable but smaller than the orbital timescale, the stellar debris of a $\sim 0.3\,M_{\odot}$ star is stretched $1.2/0.19\sim 6.3$ times during one orbital period. Since the original star covered only about $\sim 2R_{\star}/(2 \pi r)\sim 1/469$ of the orbital circumference, it is expected to take $\sim 469/6.3\sim 74$ orbital periods or $\sim 0.01$ years to stretch the debris so that they cover the whole orbit. This is in very close agreement with the results of the fits of the optical/UV light curves with 2 accretion events. 

This implies that the first peak in the light curve is likely related to the disintegration of the stellar body and related shocks with the surrounding accretion disc. The second peak is plausibly related to the filling of the original circularized stellar orbit by the stellar debris, and the whole stream thus could have started colliding with itself, which resulted in additional shocks producing enhanced optical/UV emission. Naturally, this is just a simplified picture and the whole complexity of such an event will need to be investigated in detailed magnetohydrodynamic simulations involving a dense accretion disc. In particular, since the star is not expected to be fully embedded within the accretion disc close to its tidal radius (see Fig.~\ref{fig_HR}), the hot debris will preferentially expand above and below the disc plane. Due to the fast expansion, the ejected material cools down and can form an optically thick screen that obscures the innermost regions and effectively reprocesses higher-energy radiation, see Fig.~\ref{fig_sketch_TDE} for an illustration. Furthermore, for the embedded star on a retrograde orbit, it has a negative angular momentum with respect to the gas. The strong head-on interaction with the disc will cause the angular-momentum increase, and the star will move on a more and more eccentric, nearly radial orbit as it approaches the zero angular-momentum boundary. After that, it can move on a prograde circularized orbit at a smaller radius. This hints at several peculiarities of a TDE on an initially nearly circular, retrograde orbit that is additionally partially embedded within the dense innermost parts of an AGN accretion disc.

\section{Discussion}
\label{sec:discussion}

In the previous sections, we performed a spectroscopic analysis of the \objname\ host, which confirmed that prior to the event, it was of an NLS1 type. During the pre-outburst epoch, the distance of the BLR was $\sim 3.2\times 10^4-10^5\,R_{\rm Schw}$. Modeling of the nuclear flare light curve with the viscously spreading ring constrained the involved mass to $\sim 0.3-0.4\,M_{\odot}$, with the preference for at least two accretion events separated by $\sim 0.01$ years. The analysis of the broadband optical-UV SED showed that the emission originates from the reprocessing layer at $\sim 1.6 \times 10^4\,R_{\rm Schw}$ with the characteristic temperature of $\sim 8\,900\,{\rm K}$. The BLR radius during the outburst shrinks to a comparable distance, hence the two structures may be related. The reprocessing layer hides the inner accretion disc, and hence also the TDE event characterized by a higher temperature.

In this Section, we discuss peculiarities of TDEs in AGN (Subsec.~\ref{subsec_TDE_AGN}), the numerical simulations of TDEs in AGN (Subsec.~\ref{subsec_simulations}), and the coevolution of AGN and dense nuclear star clusters (Subsec.~\ref{subsec_coevolution}), which supply stars that interact with the accretion flow surrounding SMBHs and eventually get disrupted. We mention some of the future prospects for studying TDEs in AGN in Subsection~\ref{subsec_future_prospects}.

\subsection{TDE in AGN}
\label{subsec_TDE_AGN}

TDEs in AGN are expected. One of the elements of AGN, sometimes forgotten, is the central nuclear stellar cluster \citep[NSC;][]{2009A&A...502...91S,2020A&ARv..28....4N}, which has a comparable size to globular clusters but is typically more massive. Stars in the NSC are expected to interact with the accretion disc surrounding the SMBH \citep[e.g.][]{zurek1994,vilkoviskij2002,2021dittmann}, and they may even partially form within its outer self-gravitating parts \citep[e.g.][]{collin1999,thompson2005}. Stellar collisions in the NSC can enhance the TDE rate by placing stars on highly eccentric orbits \citep{2025rose}.

Observationally, TDEs are predominantly found in quiescent galaxies \citep[see e.g.][for a review]{Gezari21_rev}. Finding a TDE in an already active galaxy is rare, despite theoretical predictions suggesting higher disruption rates in AGN than in quiescent galaxies \citep{karas2007}. The observed occurrence rate therefore, remains several times below theoretical predictions \citep{2015Merloni,kaur2025}, likely because TDE signatures are difficult to recognize against the variable AGN background, particularly for larger black hole masses \citep[e.g.][]{zhang2023}. 


\citet{Liu2020} detected the TDE candidate in SDSS
J022700.77-042020.6. Its AGN nature was confirmed by the X-ray spectral properties and the detection of
the broad optical emission lines after the flare phase. The outburst took place in 2009, and by 2016, the source had returned 
almost to its pre-flare level. The X-ray slope of $\Gamma = 2.38$ measured well before and well after the flare (years 2002 and 2017, respectively) suggests that this is also an NLS1. \citet{Liu2020} observed a plateau in the light curve, which they interpreted as due to the collision of the debris stream with a pre-existing accretion disc. This source is thus very similar in its properties to \objname.

For \objname\ the latest ATLAS light curve in both the $o$ and $c$ bands, inspected on 2025 November 2 (MJD = 60981), shows that the flux in the most recent bin (MJD = 60831–60981) is $\sim5\%$ higher than the pre-flare flux measured between MJD = 57300 and 57440. However, near-UV SWIFT (UM2-band 19.86 mag) data from 2025-11-16 compared to GALEX (NUV 20.93 mag) archival photometry, show that the source is still one magnitude brighter.

Overall, the list of well-confirmed TDE detections in AGN is still rather limited.
Besides \objname\ and SDSS J022700.77-042020.6, there is one more well-confirmed example, CSS100217 \citep{2011drake}, which also took place in an NLS1 galaxy, although its interpretation required considerable debate before a consensus was reached \citep[see][and the references therein]{gu2025}. This last event is much more massive, both in the aspect of the central black hole and the disrupted star. 
There are several other systems where a TDE-in-AGN interpretation could be considered, albeit with somewhat lower confidence. These include PS1-10adi, where the main interpretation was that of a peculiar kind of supernova \citep{2017kankare}, which would belong to the more general class of supernovae interacting with their denser environment \citep{2020A&A...642A.214K,2026arXiv260115428K}; five of the transients hosted in NLS1s, studied by \cite{2021ApJ...920...56F}, with three of them belonging to the emerging class of Bowen fluorescence flares \cite{2019Trakhtenbrot}) and two that are interpreted as TDEs in NLS1; or the peculiar CLAGN 1ES 1927+654 \citep{Trakhtenbrot2019_1ES}, which could be associated with a TDE stream interacting with a pre-existing accretion disc \citep{Ricci2020_1ES}.

There are also nuclear transients whose nature remains ambiguous, including ANTs that may reflect either changing-look AGN activity or TDEs, and more extreme disruptions that typically occur in non-active (or only weakly active) hosts \citep[see e.g.][]{2025wiseman, clark2025, hinkle2025, onori2022, hoogendam2024}. Some of the sources were not well observed before the outburst, so it is not possible to estimate their activity level before the disruption \citep[e.g.][for the source AT2018fyk]{wevers2021}. The evolution of AT2018fyk is very different from \objname, as it shows clearly the emission from the inner disc, and subsequent rebrightenings from the following passages of the star on an elliptical orbit \citep[e.g.][]{wevers2021,pasham2024}.

\subsection{Numerical simulations of TDEs in AGN environments}
\label{subsec_simulations}

Simulations of such events were pioneered by \citet{chan2019}. They showed that the disruption process is unaffected by the presence of the disc. However, the interaction between the bound debris stream and the pre-existing disc is a complex process that can strongly modify the observational appearance. In particular, the stream–disc collision can generate strong shocks, produce hot plasma, and drive highly enhanced dissipation, potentially leading to transiently super-Eddington mass inflow rates.
The subsequent papers discussed several aspects expected in TDE in AGN, as the delayed X-ray afterglow from wind collision with the dusty torus \citep{mou2021,2021zhuang}, the development of eccentricity in the disc as a result of the event \citep{weavers2022}, and even the eccentricity pumping in the discs of active galactic nuclei as a result of TDE from three-body scatterings \citep{prasad2024}. 

Most simulations focused on stars moving along highly eccentric orbits, characteristic of stars that come directly from the NSC. In this case, the evolution is initially determined by the orbital dynamics of the debris, with viscous spreading of the newly formed disc becoming important only at later times \citep[][]{rees1988,shen2014}. But if stars were first trapped inside the disc \citep[e.g.][]{1991MNRAS.250..505S}, the process proceeds differently. Since stars are not massive enough to open a gap in the disc \citep[see e.g.][]{stolc2023}, their presence before the disruption does not modify the accretion disc state considerably. In-plane tidal disruption of stars in AGN discs was studied by \citet{2024ryu}, who analyzed the evolution of the debris from main-sequence stars disrupted by a $10^6\Msun$ supermassive black hole, surrounded by a gaseous disc. 
They performed simulations for stellar orbits with different orientations relative to the disc gas, including both prograde and retrograde configurations.

They estimated the critical density of the disc above which a fully muffled TDE takes place. The critical disc density depends on several parameters: central black hole mass, orientation of the stellar orbit relative to the disc rotation, viscosity parameter, accretion rate, and radius. The dependence of the gas number density on dimensionless accretion rate is quite strong, $n \propto {\dot m}^{-2}$ \citep{ss73}. They predict that the fallback rate is super-Eddington, but it depends on the stellar mass and is clearly lower for less massive stars. 

In the source \objname\, we do not directly observe any hot plasma. This may mean that the stellar passage was not along the highly eccentric orbit, as usually assumed in TDE, but the star was already embedded in the accretion disc before disruption, as in the scenario of \citet{2024ryu}. It may also imply that we underestimate the bolometric luminosity of the \objname\ by considering the visible reprocessed radiation.

Our model assumes that the matter is deposited in a circular orbit within the accretion disc. This motivates a comparison with the scenario studied by \citet{linial2024}, in which a star orbits around an SMBH on a circular orbit. In their model, the star gradually loses mass, forming a disc from its own material before eventually undergoing tidal disruption. This produces an asymmetric light curve, with a shallower decay than the rise in luminosity. The characteristic post-disruption plasma temperature is estimated to be $10^5$--$10^6$ K, implying that the emission region would need to be hidden within an envelope to reduce the effective temperature to  $\sim 8,900$~K observed in \objname. However, the evolution would differ if the star were embedded in the accretion disc from the beginning, as it could instead accrete mass from the disc rather than lose it.






\subsection{Co-evolution of an AGN and the NSC}
\label{subsec_coevolution}

 The presence of the stellar population within the inner part of the NSC is relatively well studied in the case of the Milky Way, due to its proximity. However, it is expected to be present in most typical galaxies \citep{2020A&ARv..28....4N}. The stellar orbits in this region are expected to be inclined relative to the accretion disc plane, causing the stars to gradually sink towards the center due to hydrodynamical friction. This raises the question of the physical properties and feeding of the central SMBH and whether a better understanding of TDEs will aid in probing the AGN environment \citep{2016MNRAS.460..240K,2025arXiv250720232H}.

The mutual interaction between stellar bodies and the gas and dust surrounding AGN induces the secular evolution of stellar trajectories. This evolution reflects the physical properties of galactic cores. While a close passage of a star near an SMBH may damage the star, the gradual removal of its orbital energy and angular momentum can increase the likelihood of its capture without being immediately destroyed \citep{1996ApJ...470..237A,2001A&A...376..686K}. When embedded in the orbital plane of the central object, the fate of the star is governed by the ratio of its radius of influence versus the accretion disc scale-height. It loses the angular momentum gradually and spirals down to the centre \citep{1994ApJ...423..581A}. Before these orbiting stars plunge into the SMBH, they are expected to contribute to the generation of gravitational waves in the Extreme Mass Ratio Inspirals regime. This additional contribution could potentially redefine future gravitational wave experiments by contributing to the background signal \citep{2000ApJ...536..663N,2024ApJ...973..101L,2024arXiv241012090Z}.

For the orbits inclined with respect to the disc plane, the overall effect on the accretion flow structure is determined by the statistics of impacts on the accretion disc. This, in turn, depends on the distribution of orbiting stars within the inner cluster and its gradual evolution \citep{1994MNRAS.267..557P,1995MNRAS.275..628R,2025ApJ...988L..21R}. Furthermore, the perturbed distribution of stars orbiting close to SMBHs (on sub-parsec scales) deviates from the general sphericity and isotropy of an NSC \citep{2004MNRAS.354.1177S}. This deviation is more pronounced in the inner regions, where the stellar population becomes flattened to the disc plane; in this scenario, the hydrodynamical collisions between the orbiter and the accretion flow generally lead to corotation. In contrast, the outer region of the cluster (on parsec scales and further out) is almost spherical and isotropic, as it eventually connects with the galactic bulge.

Concerning counter-orbiting stars, given the low inferred mass of $\sim 0.3-0.4\,M_{\odot}$ for \objname, they are expected to be less abundant than coorbiting stars that are more natural, given the alignment by the hydrodynamical drag of the accretion disc. Essentially all inclined stars interacting with the disc are expected to align with it on the alignment timescale, see Eq.~\eqref{eq_alignment_timescale} and Fig.~\ref{fig_timescales}. Counter-orbiting stars can enter the disc within its plane by dynamical relaxation processes, such as resonant and two-body (non-resonant) relaxation. Once they enter at the outer disc edge, the strong hydrodynamical drag due to the head-on interaction can rapidly decrease the stellar orbital radius within $\lesssim 25\,000$ years (see Fig.~\ref{fig_timescales}). During this time, the orbital angular momentum of the star (negative with respect to the disc) increases, and at some point the stellar orbit becomes nearly radial--plunging. The star is then expected to become corotating with the disc at a much smaller radius. At this time, it would be significantly ablated and likely close to its tidal radius. Another mechanism by which a stellar orbit could be flipped from one sense to another, even within a single stellar disc formed within a gaseous disc, is the Kozai-Lidov (KL) oscillation due to a massive perturbing body, such as a massive stellar/gaseous disc itself or another massive black hole \citep{2015MNRAS.451.1341L,2016ARA&A..54..441N}. The basic condition for the KL eccentricity-inclination oscillation is a larger inclination $\iota \gtrsim 40^{\circ}$. However, a dedicated numerical integration is required to verify the applicability of the KL oscillation under the combined action of both a massive perturber and an AGN accretion disc, which is expected to swiftly decrease the inclination offset.

\subsection{Future prospects}
\label{subsec_future_prospects}

The interpretation of the outburst in \objname\ as a TDE partially hidden within a highly temporally obscured AGN and the subsequent evolution of a star on a roughly circular counter-rotating orbit motivates future searches for other NLS1s that may exhibit similar episodes of transient activity characterized by an enhanced optical or UV flux and a temporary reduction in the X-ray band. Stars interacting with the disc are considered to be attractive sources of gravitational radiation for the LISA mission, and in general an attractive topic for multimessenger astronomy \citep[e.g.][]{kejriwal2024,2025ApJ...987L..11O}. On-going and future all-sky surveys performed by e.g.  Vera C. Rubin Observatory’s Legacy Survey of Space and Time \citep{2019ApJ...873..111I}, ULTRASAT \citep{2024Shvartzvald}, and the subsequent high-cadence follow-up observations \citep[e.g. with QUVIK;][]{2022arXiv220705485W,2024SSRv..220...11W,2024SSRv..220...24K,2024SSRv..220...29Z,2025JATIS..11d2222Z} should bring an unprecedented possibility to find many events, and to build a statistics of their properties. 

   

\section{Conclusions}
\label{sec:conclusions}

The main results concerning the modelling of the \objname\ event associated with the nucleus of the Narrow Line Seyfert 1 (NLS1) galaxy SDSS J015804.75-005221.8 ($z=0.080440$) may be summarized as follows:
\begin{itemize}
    \item The modelling of the accretion event light curve, including viscous spreading, constrains the disrupted mass to $\sim 0.3-0.4\,M_{\odot}$.
    \item For a $\sim 0.3\,M_{\odot}$ star the tidal disruption would take place at $\sim 13$ Schwarzschild radii of a $\sim 10^6\,M_{\odot}$ supermassive black hole.
    \item There appears to be a preference for multiple, at least two accretion events separated by $\sim 4$ days.
    \item The UV-optical SED is consistent with the combination of a single black-body ($8\,900\,{\rm K}$) and BLR emission. Given the bolometric luminosity of the event of $\sim 10^{44}\,{\rm erg\,s^{-1}}$, the UV-optical SED is consistent with an optically thick reprocessor at $\sim 10^4$ Schwarzschild radii, which is likely related to the outflow following the TDE.
    \item Dynamical estimates indicate that the TDE could have resulted from the star orbiting on a circular orbit within the accretion disc. A low mass of $\sim 0.3\,M_{\odot}$ is more consistent with a counter-orbiting star as the coorbiting star would gain additional mass, eventually exceeding $\sim 1\,M_{\odot}$ during the inspiral.
    \item The star moving in an opposite sense to the disc quickly spiraled inwards within $r\sim 1000\,R_{\rm Schw}$ on a timescale of $\lesssim 100$ years. During this process, it also underwent rapid ablation due to ram pressure, therefore it was initially more massive ($\sim 0.6$--$2.1\,M_{\odot}$).
    \item At the tidal radius of $\sim 10\,R_{\rm Schw}$, a $\sim 0.3\,M_{\odot}$ main-sequence star is geometrically thicker than a standard accretion disc, which can lead to the preferential fast expansion of the debris above and below the disc plane, eventually creating an optically thick reprocessing layer at larger distances. Multiple accretion events can be related to the tidal spreading of the debris along the eccentric orbit and subsequent interaction with the surrounding disc and itself. 
\end{itemize}

\begin{acknowledgements}
We are grateful to the Referee for their helpful and constructive remarks which helped to improve considerably the manuscript.
MS acknowledges the Czech Science Foundation (GA\v{C}R) grant no. 26-23342I. 
This project has received funding from the European Research Council (ERC) under the European Union’s Horizon 2020 research and innovation program (grant agreement No. [951549]). The Czech-Polish Mobility program of the two Academies of Sciences, titled ``Appearance and dynamics of accretion onto black holes'', is greatly appreciated. BC, VK and PK acknowledge the OPUS-LAP/GA bilateral project ``Weather effects in using disc continuum time delays in AGN to measure the expansion rate of the Universe'' (2021/43/I/ST9/01352/OPUS 22 and GF23-04053L). MZ acknowledges the financial support of the Czech Science Foundation Junior Star grant no. GM24-10599M. MS and BT acknowledge support from the European Research Council (ERC) under the European Union’s Horizon 2020 research and innovation program (grant agreement number 950533), and the Israel Science Foundation (grant number 1849/19). VK thanks the Czech Science Foundation (ref.\ 21-06825X). TJ conducts his research under the Marie Skłodowska-Curie Actions – COFUND project, which is co-funded by the European Union (Physics for Future – Grant Agreement No. 101081515). TP acknowledges the financial support from the Slovenian Research Agency (grants I0-0033, P1-0031, J1-8136, J1-2460 and N1-0344). DI acknowledges funding provided by the University of Belgrade -- Faculty of Mathematics through the grant (the contract 451-03-136/2025-03/200104) of the Ministry of Science, Technological Development and Innovation of the Republic of Serbia.
\end{acknowledgements}

%
%
\bibliographystyle{aa}
\bibliography{aa}

\end{document}